\documentclass[showpacs,pre,amssymb,amsmath,preprint]{revtex4}

\usepackage{amsmath}
\usepackage{amssymb}
\usepackage[utf8]{inputenc}
\usepackage{InstructionDiagramDavid}

\newcommand{\dd}{\hbox{d}}
\newcommand{\br}{\mathbf{r}}
\newcommand{\bp}{\mathbf{p}}

\newcommand{\bk}{\mathbf{k}}

\newcommand{\be}{\begin{equation}}
\def \la{\label}

\begin{document}

\title{Fourth moment sum rule for the charge correlations of a two-component classical plasma}


\author{Angel Alastuey $\dag$ and Riccardo Fantoni $\ddag$}


\affiliation{ $\dag$ Laboratoire de Physique de l'Ecole Normale Sup\'erieure de Lyon, Universit\'e de Lyon and CNRS, 46, all\'ee d'Italie, F-69007 Lyon, France}

\affiliation{ $\ddag$ Universit\`a di Trieste, Dipartimento di Fisica, strada
Costiera 11, 34151 Grignano (Trieste), Italy}

\date{Version: \today}

\begin{abstract}

We consider an ionic fluid made with two species of mobile particles carrying 
either a positive or a negative charge. We derive a sum rule for the fourth moment of equilibrium 
charge correlations. Our method relies on the study of the system response 
to the potential created by a weak external charge distribution with 
slow spatial variations. The induced particle densities, and the resulting induced charge density, 
are then computed within density functional theory, where the free energy is expanded 
in powers of the density gradients.
The comparison with the predictions of linear response theory provides a thermodynamical 
expression for the fourth moment of charge correlations, which involves the isothermal 
compressibility as well as suitably defined 
partial compressibilities. The familiar Stillinger-Lovett condition is also recovered as a by-product
of our method, suggesting that the fourth moment sum rule should hold in any 
conducting phase. This is explicitly checked in the low density regime, within the Abe-Meeron 
diagrammatical expansions. Beyond its own interest, the fourth-moment sum rule should be useful for 
both analyzing and understanding recently observed behaviours near the ionic critical point. 
 
\end{abstract}


\maketitle

\section{Introduction}

Sum rules have been playing an important role in the study of charged systems for many years. In 
general, a sum rule provides a relation between microscopic correlations on the one hand, and macroscopic 
or universal quantities on the other hand. For charged systems, sum rules often express screening properties, so 
they shed light on the fundamental mechanisms at work. Furthermore, they also provide useful constraints 
for approximate theories. Sum rules have been derived for a large variety of charged systems, including classical, quantum and 
relativistic plasmas, while they concern both static and dynamic properties in the bulk or near interfaces. 
For instance, let us mention the last work~\cite{Janco2010} by Bernard Jancovici, devoted to the study of the time-displaced 
charge correlations of a relativistic one-component plasma coupled to radiation. Other examples, 
including many contributions from Jancovici, can be found in 
two reviews~\cite{Martin1988,BryMar1999}. 

One of the most well-known sum rules for classical ionic fluids was derived long ago by 
Stillinger and Lovett~\cite{StiLov}, who shown that the second moment of equilibrium charge 
correlations is given by a simple universal expression, valid in any plasma phase
and independent of the microscopic details of the considered models. That second-moment sum 
rule expresses the perfect screening of weak external charges. A few years later, Vieillefosse and Hansen~\cite{Vieille} 
derived another sum rule for the fourth moment of the charges correlations of the one-component plasma (OCP), where 
identical positively charged particles move in a rigid uniform neutralizing background. That fourth moment is 
expressed in terms of the isothermal compressibility. Soon after that work, 
there was an attempt~\cite{Gia} to extend such a fourth-moment sum rule to the two-component plasma (TCP) where both positive and 
negative charges are mobile. The corresponding expression for the fourth moment involves 
ill-defined thermodynamic quantities, so its validity remained quite doubtful. A more convincing approach for that TCP 
was introduced by 
van Beijeren and Felderhof~\cite{vBF1979}. However, the thermodynamical quantities involved in the expression 
of the fourth moment are defined through the application of suitable external potentials which are not explicited, 
while the derivation itself is rather tough. In fact, a similar expression was obtained later by 
Suttorp and van Wonderen~\cite{Suttorp1987} for a multicomponent ionic mixture (MIM), where 
all mobile charges have the same sign and interact 
\textsl{via} the pure Coulomb potential, while a rigid uniform background of opposite charge ensures overall neutrality. Then, 
all involved thermodynamic quantities become well defined within the considered MIM.

The main goal of the present paper is to derive a fourth moment sum rule for a general TCP, namely to express such moment 
in terms of suitably defined thermodynamical quantities, similarly to the formulae derived for the OCP~\cite{Vieille} 
or the MIM~\cite{Suttorp1987}, \cite{Suttorp2008}. Our strategy, inspired by Jancovici's style, consists in studying the 
response of the TCP to a weak external charge distribution with a plane wave structure. In the long wavelength limit, the 
induced local particle densities vary on macroscopic scales. This allows us to compute the response within some 
hydrostatic-like approach which involves local equilibrium states with arbitrary densities. As a crucial point, a 
proper definition of equilibrium homogeneous non-neutral states with arbitrary densities naturally emerges  
within the framework of density functional theory (DFT). Then the induced charge density is expressed in terms of 
well-behaved thermodynamical quantities of an auxiliary system, which is nothing but a TCP immersed in 
a rigid uniform neutralizing background. Comparing that expression with the general linear response formula, we obtain the 
required fourth moment sum rule for the charge correlations of the genuine unperturbed TCP. 

According to the previous strategy, we first introduce in Section~\ref{SI} the various systems which intervene in 
our analysis. Of course, we start by defining the TCP, where a short-range regularization of the pure Coulomb 
interactions is essential for avoiding the classical collapse between oppositely charged particles. Two 
examples of such regularizations are provided, associated with either soft or hard spheres. After recalling that 
the TCP is always neutral and homogeneous in the absence of any external action on the particles, 
we show how the application of a suitable external potential produces homogeneous non-neutral states. This 
leads to the introduction of an auxiliary system, the TCP immersed in a charged background, for which 
equilibrium states are well defined for any set of particle densities. 

The general framework of DFT is exposed in Section~\ref{DFT}, where we provide the fundamental DFT equation 
which relates particle densities to the applied external potentials. The central object in that relation is the 
free energy, which is a functional of particle densities. For slow spatial variations, that free energy 
functional can be expanded in powers of the gradients of particle densities, where the local ingredients 
are equilibrium quantities of the above homogeneous auxiliary system. Let us mention 
that the idea of using density-gradient expansions was 
introduced a long time ago by van der Waals~\cite{vdW1894} for studying capillarity.  

In Section~\ref{LR}, within DFT, we compute the induced particle densities of the TCP submitted to a weak 
external charge distribution with a plane wave structure and wavenumber $k$. 
The resulting induced charge density exactly cancels the 
external charge distribution in the long wawelength limit $k \to 0$, as expected from perfect screening arguments. 
Furthermore, its amplitude at the order $k^2$ included only depends on thermodynamical quantities of 
the auxiliary system. In other words the square-gradient corrections in the free energy functional, 
which intervene in the corresponding amplitudes of each induced particle density at this order, do 
not contribute anymore when forming the charge density thanks to cancellations. Then, by comparing this 
exact expression of the induced charge density obtained by DFT with the linear response formula, we obtain 
the required sum rule for the fourth moment of the charges correlations of the homogeneous neutral TCP. 
The corresponding thermodynamical expression of that fourth moment involves not only the isothermal 
compressibility of the TCP, but also partial compressibilities specific to the auxiliary system. We 
briefly discuss the content of previous approaches~\cite{Gia}, \cite{vBF1979}, and we show how the 
known results for the OCP~\cite{Vieille} and the MIM~\cite{Suttorp2008} can be easily recovered within our general method. 

It is worthy to check explicitly the fourth moment sum rule for specific models  
where exact calculations can be carried out for both microscopic and thermodynamical quantities. 
In Section~\ref{LDE}, we consider a model of charged soft spheres in the low density limit at fixed 
temperature. Within the Abe-Meeron resummations of the familiar Mayer diagrammatics for particle correlations, 
we first compute the lowest order terms in the density expansion of the fourth moment of charge correlations, namely the 
terms of order $1/\rho$, $1/\rho^{1/2}$, $\rho \ln \rho$ and $\rho^0$ in the density $\rho$. Abe-Meeron 
resummed diagrammatics also provide the low density expansion of the thermodynamical quantities involved in the 
fourth moment sum rule : the corresponding expansion of the thermodynamical expression of the fourth moment 
exactly coincides with the previous purely microscopic calculation up to order $\rho^0$ included. That remarkable agreement holds 
for any values of the microscopic parameters defining the model.

In Section~\ref{Conclusion}, we provide some additional comments about the derivation itself and its underlying assumptions, 
as well as extensions to three and more component systems. Beyond its own conceptual interest, we also discuss a
possible use of the fourth moment sum rule for a better understanding of the conductor or dielectric nature of the 
critical point of the liquid-gas transition of an ionic fluid. It turns out that recent 
Monte Carlo simulations~\cite{DasKimFisher2011} 
strongly suggest that the fourth moment of charge correlations diverge when approaching the critical point, in 
a way close to that of the isothermal compressibility. That observation was one of the motivations for the present work.

\section{The systems of interest}
\la{SI}

\subsection{Examples of two-component plasmas}

We consider a two-component classical plasma (TCP) made with two species $\alpha=1,2$ of mobile particles carrying 
positive or negative charges, let us say $q_1 =Z_1 q > 0$ and $q_2=-Z_2 q < 0$ with $Z_1$ and 
$Z_2$ positive integers. 
The corresponding Hamiltonian for a total number of particles $N=N_1 + N_2$ reads 
\be
\label{HTCP}
H_{N_1,N_2}= \sum_{i=1}^N \frac{\bp_i^2}{2m_{\alpha_i}} + 
\frac{1}{2} \sum_{i \neq j} u_{\alpha_i\alpha_j}(\br_i,\br_j)
\end{equation}
where $\alpha_i=1,2$ is the species of particle i.
The two-body potential $u_{\alpha_i\alpha_j}(\br_i,\br_j)$ only depends on the relative distance
$r=|\br_i-\br_j|$, and it includes some short-range regularization of the 
Coulomb interaction, which is crucial for avoiding the classical collapse between oppositely charged 
particles. A first soft regularization is embedded in the simple expression 
\begin{equation}
\label{soft}
u_{\alpha\gamma}(r)=\frac{q_{\alpha}q_{\gamma}}{r} [1-\exp(-r/d_{\alpha\gamma})] \; 
\end{equation}
which remains finite at $r=0$. The lengths $d_{\alpha\gamma}$ 
control the exponential decay at large distances of the short-range part of the 
two-body potential. 

A second regularization amounts 
to introduce hard cores, namely
\be
\la{2body}
u_{\alpha\gamma}(r) = \infty  \; \text{for} \; r < \sigma_{\alpha\gamma} 
\; \text{and} \; u_{\alpha\gamma}(r) = \frac{q_\alpha q_\gamma }{r} \; \text{for} \; r > \sigma_{\alpha\gamma} \;. 
\end{equation}
The corresponding TCP of charged hard spheres is suitable for describing many ionic mixtures, where the 
hard-core interaction mimics the effective repulsion between the electronic clouds of two ions. If 
$\sigma_{11}$ and $\sigma_{22}$ can be viewed as the effective diameters of the ions, the characteristic 
crossed lengths $\sigma_{12}=\sigma_{21}$ differ in general from the average $(\sigma_{11} +\sigma_{22})/2$ which 
would arise if particles really were billiard balls. This so-called non-additivity can be understood by noticing 
that the $\sigma_{\alpha\gamma}$'s are the typical ranges of the repulsions between electronic clouds 
for which no pure geometrical considerations apply. The simplest version of that general asymmetric TCP 
is the celebrated Restrictive Primitive Model, which is fully symmetric with respect to the 
charges and the hard-core diameters, namely
$|q_1|=|q_2| = q$ and $\sigma_{11}=\sigma_{22}=\sigma_{12}=\sigma$. 

Other short-range regularizations of the Coulomb interaction can be introduced. The corresponding most general 
TCP will be denoted ${\cal S}$. The following derivations are valid for any ${\cal S}$,  
beyond the above two examples.

\subsection{The homogeneous neutral TCP}

Let us first consider that ${\cal S}$ is enclosed in a box with volume $\Lambda$, while no 
external potential is applied to the particles. At equilibrium, 
all statistical ensembles should become equivalent in the thermodynamic limit which is also assumed to exist. 
Furthermore, in a fluid phase, the bulk is overall neutral, that is the 
homogeneous particle densities $\rho_1$ and $\rho_2$ far from the boundaries satisfy 
the local neutrality relation
\be
\la{neutrality}
q_1 \rho_1 + q_2 \rho_2 =0   \;. 
\end{equation}
Strictly speaking, these remarkable results have been only proved in the Debye regime, namely 
at sufficiently high temperatures and sufficiently low densities, for rather general 
regularized interactions and rational ratios $q_2/q_1$~\cite{BryFed1980}. Moreover, there exists 
a proof for charge symmetric systems,
\textit{i.e.} $q_{1}=-q_{2}$, for any values of the thermodynamic parameters~\cite{FroPar1978}. Let us also
mention the beautiful proof for the
quantum version with pure Coulomb interactions by Lieb and Lebowitz~\cite{LL72}. According to all 
those rigorous results, it can be reasonably expected that both the existence of the thermodynamic limit 
and the local neutrality are valid for any classical TCP in the whole fluid phase.  

Important features of the various statistical ensembles are associated with the neutrality 
relation~(\ref{neutrality}). In the grand-canonical ensemble, the intensive thermodynamical parameters 
are the inverse temperature $\beta$ and the chemical potentials $\mu_\alpha$. It turns out that only 
the linear combination $\mu =(Z_2\mu_1 + Z_1 \mu_2)/(Z_1+Z_2)$ is relevant and entirely determines 
the total particle density $\rho=\rho_1 + \rho_2$. This can be readily understood within the following 
simple heuristic arguments. Let us introduce, 
for any arbitrary configuration, the total number 
of particles $N=N_1+N_2$ and the corresponding total charge $Q=Mq$ with $M=Z_1 N_1-Z_2 N_2$.  According to the 
decomposition
\be
\la{muNdecomposition}
\mu_1 N_1 + \mu_2 N_2= \mu N + \nu M  \; 
\end{equation}
with $\nu=(\mu_1-\mu_2)/(Z_1+Z_2)$, we see that $\mu$ controls the grand-canonical average $<N>_{\rm GC}$ 
of the total particle number, while $\nu$ determines the  grand-canonical average $<Q>_{\rm GC}=q<M>_{\rm GC}$  
of the net charge. In the thermodynamic limit (TL), $\Lambda \to \infty$ with $\beta$ and $\mu_{\alpha}$ fixed, 
the contributions of non-neutral configurations with $Q$ proportional to the volume $\Lambda$ become 
negligible, because the corresponding Boltzmann factors involve a positive self-electrostatic energy 
which diverges faster than $\Lambda$ itself. Accordingly, the total charge density in the bulk 
\be
\la{GCneutrality}
q_1 \rho_1 + q_2 \rho_2=\lim_{\rm TL}<Q>_{\rm GC}/\Lambda \; 
\end{equation}
vanishes for any given $\nu$, while the total particle density 
$\rho=\rho_1 + \rho_2$ is indeed entirely determined by $\mu$ and $\beta$. 

In the canonical ensemble, the TL is defined by letting $\Lambda \to \infty$ and $N_\alpha \to \infty$, keeping 
$\beta$ and $N_\alpha/\Lambda$ fixed. All excess charges go to the the boundaries in the TL, and the remaining bulk is always 
neutral. The bulk thermodynamic quantities and bulk distribution functions computed within the canonical 
ensemble then become identical to their grand canonical counterparts. In particular the free-energy density in thermal units
of this homogeneous neutral phase, which only 
depends on $\rho$ and $\beta= 1/(k_{\rm B}T)$, can be computed through
\be
\la{FEneutral}
f(\rho,\beta) = \lim_{\rm TL}   (\beta \sum_\alpha \mu_\alpha<N_\alpha>_{\rm GC}  -  \ln \Xi)/\Lambda \; ,
\end{equation}
where $\Xi$ is the grand-canonical partition function. This provides the familiar thermodynamic identity 
\be
\la{FEneutraldensity}
f(\rho,\beta) =  \beta (\rho \mu  - P) \; ,
\end{equation}
with the pressure $P= \lim_{\rm TL} k_{\rm B}T \Lambda^{-1}\ln \Xi $. Since the pressure is also given by 
the thermodynamic relation 
\be
\la{pressure}
\beta P  =  \rho \frac{ \partial f}{\partial \rho  }   
- f(\rho,\beta) \; ,
\end{equation}
we infer
\be
\la{ChePotNeutral}
\beta  \mu  = \frac{ \partial f}{\partial \rho  }   \; , 
\end{equation}
which is analogous to the standard thermodynamical identity expressing the chemical potential 
for a one-component system with short-range interactions.

\subsection{The homogeneous non-neutral TCP in an external potential}

In order to obtain a non-neutral homogeneous state of ${\cal S}$ with arbitrary uniform densities, 
one must apply a non-vanishing external potential on 
the particles. Let us introduce the electrostatic 
potential $\varphi_{\rm B}(\br)$ created by an uniform charge density $c_{\rm B}$, 
and the corresponding external potentials $U_\alpha^{{\rm B}}(\br)= q_\alpha \varphi_{\rm B}(\br)$ 
seen by the particles. At equilibrium, the total electrostatic field inside the 
bulk should identically vanish. According to that simple electrostatic argument, 
the induced particle densities 
should be homogeneous, while the resulting charge density $q_1 \rho_1 + q_2 \rho_2$ 
carried by the particles should cancel the external charge density $c_{\rm B}$.

Interestingly, the above quite plausible scenario has been exactly demonstrated 
within a solvable model by Jancovici~\cite{Janco81}. He considered identical point particles in two dimensions
with pure Coulomb interactions, which then take a logarithmic form. In addition the particles are submitted 
to a confining parabolic potential, associated with a fixed external uniform charge density. For a 
special value of the temperature, all equilibrium distribution functions can be exactly 
computed. The resulting particle density is indeed uniform in the bulk and such that the total charge density 
vanishes. Furthermore, all higher-order distribution functions in the bulk become 
translationally invariant in the TL.

\subsection{The auxiliary system in a neutralizing rigid background}

As suggested by the previous considerations, and for further purposes, 
it is convenient to introduce an auxiliary system ${\cal S}^\ast$, where now the mobile 
positive and negative charges of the TCP are immersed in an uniform rigid background with charge 
density $c_{\rm B}$. The corresponding Hamiltonian of ${\cal S}^\ast$ reads
\be
\la{HTCPB}
H_{N_1,N_2}^\ast= \sum_{i=1}^N \frac{\bp_i^2}{2m_{\alpha_i}} + 
\frac{1}{2} \sum_{i \neq j} u_{\alpha_i\alpha_j}(\br_i,\br_j) + 
\sum_{i=1}^N \int_\Lambda \dd \br  \frac{q_{\alpha_i} c_{\rm B} }{|\br_i-\br|} 
+ \frac{1}{2} \int_{\Lambda^2} \dd \br \dd \br' \frac{c_{\rm B}^2 }{|\br'-\br|} \; ,
\end{equation}
when the system is enclosed in a box with volume $\Lambda$. That system can be viewed as 
an extension of the well-known One-Component Plasma (OCP) made of identical charged particles immersed 
in a neutralizing rigid background. Now, there are two species which are immersed in the background, 
similarly to the case of a Binary Ionic Mixture (BIM). However, notice that here we do need 
a short-range regularization of the Coulomb interaction in order to avoid 
the collapse between oppositely charged particles, while the BIM can be defined with pure $1/r$ Coulomb interactions
because all mobile charges have the same sign.

Like the OCP or the BIM, the system ${\cal S}^\ast$
should have a well-behaved thermodynamic limit, 
which is now taken with a fixed background charge density $c_{\rm B}$. Now, in the bulk region, 
which is again electrically neutral, the homogeneous particle densities satisfy the neutrality relation
\be
\la{neutralitybis}
 q_1\rho_1 + q_2 \rho_2 + c_{\rm B}=0 \;.  
\end{equation} 
The corresponding free-energy density $f^\ast$ of the homogeneous neutral system now depends on $\beta$, 
$c_{\rm B}$ and one particle density. Equivalently, $f^\ast$ depends on $\beta$ and on the two particle densities 
$\rho_1$ and $\rho_2$. For any given set $(\rho_1,\rho_2)$,
the charge background density is adjusted in order to satisfy the neutrality relation~(\ref{neutralitybis}).
This defines the function $f^\ast(\rho_1,\rho_2,\beta)$, where now $\rho_1$ and $\rho_2$ are independent variables.     
That procedure is analogous to that which defines the free-energy density $f_{\rm OCP}(\rho,\beta)$ 
of the OCP for any value of the particle density $\rho$ where a suitable background charge density always ensure overall 
neutrality. 

The homogeneous neutral TCP can be viewed as a particular realization of ${\cal S}^\ast$ for 
densities $(\rho_1,\rho_2)$ satisfying the neutrality relation~(\ref{neutrality}) in the absence of any 
background. For such neutral sets, each particle density can be expressed in terms of the 
total particle number density as 
\be
\la{neutraldensities}
 \rho_1 =\frac{Z_2}{Z_1+Z_2} \rho \;\;\; \text{and} \;\;\;  \rho_2 =\frac{Z_1}{Z_1+Z_2} \rho \;.  
\end{equation} 
The thermodynamic quantities of the homogeneous neutral TCP can then be inferred from their 
counterparts of ${\cal S}^\ast$ for the neutral set of densities~(\ref{neutraldensities}).
For instance, the free-energy density of the homogeneous neutral TCP is given by
\be
\la{FEneutralbis}
f(\rho,\beta) = f^\ast(Z_2 \rho/(Z_1+Z_2), Z_1 \rho/(Z_1+Z_2),\beta) \;.  
\end{equation} 
For further purposes, it is useful to consider the isothermal compressibility defined by
\be
\la{IsoComp}
\chi_T= - \lim_{\rm TL} \Lambda^{-1} \frac{\partial \Lambda}{\partial P}= 
\left[\rho \frac{\partial P}{\partial \rho}\right]^{-1}   =  
\beta \left[\rho^2 \frac{\partial^2 f}{\partial \rho^2}\right]^{-1}   \;,  
\end{equation}
where all partial derivatives are taken at fixed $\beta$. According to identity~(\ref{FEneutralbis}), 
$\chi_T$ can be recast as 
\be
\la{IsoCompbis}
\chi_T =  
\frac{\beta (Z_1+Z_2)^2}{\rho^2 } [Z_2^2\frac{\partial^2 f^\ast}{\partial \rho_1^2}+ Z_1^2\frac{\partial^2 f^\ast}{\partial \rho_2^2}\ + 2 Z_1 Z_2\frac{\partial^2 f^\ast}{\partial \rho_1 \partial \rho_2}]^{-1}   \;,  
\end{equation}
where the second order partial derivatives of $f^\ast$ are evaluated at the neutral set~(\ref{neutraldensities}).
Eventually, all distribution functions of the homogeneous neutral TCP obviously reduce to those of ${\cal S}^\ast$ 
for that set of densities.

\section{Density functional theory}
\la{DFT}

\subsection{Grand-canonical description}

Now we consider a general inhomogeneous state of ${\cal S}$, where each particle of species $\alpha$ is 
submitted to an external potential $U_{\alpha}(\br)$. We define the inhomogeneous fugacity  
of each species by
\be
\la{fugacityalpha}
z_\alpha (\br) = \frac{\exp [\beta (\mu_\alpha -U_{\alpha}(\br)) ]}
{(2 \pi \lambda_\alpha^2)^{3/2}}\;.  
\end{equation} 
where $\lambda_\alpha= (\beta \hbar^2/m_\alpha)^{1/2}$ is the de Broglie 
thermal wavelength of species $\alpha$. The classical grand-canonical partition 
function of ${\cal S}$ enclosed in a box with volume $\Lambda$ reads
\be
\la{GCPF}
\Xi = \sum_{N_1,N_2=0}^\infty \frac{1}{N_1!N_2!}
\int \prod_{i=1}^N \dd \br_i \; z_{\alpha_i} (\br_i) \; 
\exp(-\beta V_{N_1,N_2}) \;, 
\end{equation} 
where $V_{N_1,N_2}$ is the potential part of the Hamiltonian~(\ref{HTCP}). The 
inhomogeneous particle density $\rho_\alpha (\br) $ can be expressed as a functional derivative of 
$\Xi$ with respect to $z_\alpha (\br)$, namely 
\be
\la{densityalpha}
\rho_\alpha (\br) = z_\alpha (\br) \frac{ \delta \ln \Xi}{\delta z_\alpha (\br) } 
\end{equation} 
while parameters $\Lambda$ and $\beta$ are kept fixed.
The free-energy ${\cal F}$ in thermal units of ${\cal S}$ 
is given by the Legendre transformation 
\be
\la{FE}
{\cal F}=\sum_\alpha \int_\Lambda \dd \br \; \rho_\alpha (\br)\; 
\beta (\mu_\alpha -U_{\alpha}(\br)) -\ln \Xi \; .
\end{equation}

\bigskip

The grand-partition function $\Xi$, as well as the free-energy ${\cal F}$, 
can be considered as functionals of either $z_\alpha (\br)$ or $\rho_\alpha (\br)$. The
functional derivative of ${\cal F}$ with respect to $\rho_\alpha (\br)$ is readily computed as 
\be
\la{functionalderivativeFE}
\frac{ \delta {\cal F}}{\delta \rho_\alpha (\br) } = \beta (\mu_\alpha -U_{\alpha}(\br)) 
\end{equation}
where we have used identity~(\ref{densityalpha}) as well as standard calculation rules for 
functional differentiation. The relation~(\ref{functionalderivativeFE}) will play a key role 
in the following. The density profiles  $\rho_\alpha (\br)$ for each given sets $(\mu_\alpha -U_{\alpha}(\br)) $
can be determined from that relation if one knows the functional dependence of ${\cal F}$ with respect to 
the inhomogeneous densities : this is the strategy of density functional theories (DFT). However, the 
main difficulty of DFT is that the free-energy functional is not exactly known, except for  
hard rods in one dimension~\cite{Percus}. In general, 
approximate forms are used. Here, we will use exact asymptotic expansions for 
densities with infinitely slow spatial variations.

\subsection{Homogeneous systems}

As argued in the previous Section, homogeneous states of ${\cal S}$ 
with arbitrary densities $(\rho_1,\rho_2)$ are 
obtained by applying the external potentials 
\be
\la{ExtPotUniform}
U_\alpha^{{\rm B}}(\br)= q_\alpha \varphi_{\rm B}(\br)= q_\alpha \int_\Lambda \dd \br'  \frac{c_{\rm B} }{|\br'-\br|}
\end{equation}
with the external charge density $c_{\rm B}=-(q_1\rho_1 + q_2 \rho_2)$. If we introduce the potential part 
$V_{N_1,N_2}^{\ast}$ of the Hamiltonian~(\ref{HTCPB}) for the auxiliary system ${\cal S}^\ast$ 
with background charge density $c_{\rm B}$, we can rewrite
\be
\la{RelationPE}
V_{N_1,N_2} + \sum_{i=1}^N U_{\alpha_i}^{{\rm B}} (\br_i)=
V_{N_1,N_2}^{\ast}- W_{\rm B} 
\end{equation} 
where 
\be
\la{SEEB}
W_{\rm B}= \frac{1}{2} \int_{\Lambda^2} \dd \br \; \dd \br' \frac{c_{\rm B}^2 }{|\br'-\br|} \; 
\end{equation} 
is the self-electrostatic energy of the background charge density $c_{\rm B}$. 
Inserting relation~(\ref{RelationPE}) into 
the general expression~(\ref{GCPF}), we obtain 
for the grand-partition function of ${\cal S}$ 
submitted to the external potentials $U_{\alpha}^{{\rm B}}(\br)$,  
\be
\la{GCPFidentity}
\Xi \{\mu_1 - U_{1}^{{\rm B}}(\cdot),\mu_2 - U_{2}^{{\rm B}}(\cdot)  \} = 
\Xi^{{\ast}} \{\mu_1 , \mu_2 \} \exp(\beta W_{\rm B}) \; ,
\end{equation} 
\bigskip
where $\Xi^{{\ast}}$ is the grand-partition function ${\cal S}^\ast$  for the same 
chemical potentials $\mu_\alpha$ and a background charge density $c_{\rm B}$, 
without any applied external potentials, 
\textsl{i.e.} $U_{\alpha}(\br)=0$. In the derivation of 
identity~(\ref{GCPFidentity}), we have used that 
$W_{\rm B}$ is a pure constant which does not depend on the particle degrees of freedom. 
This also implies that all the grand-canonical averages for ${\cal S}$ submitted to $U_{\alpha}^{{\rm B}}(\br)$
are identical to those for ${\cal S}^\ast$. In particular, both particle densities are identical, so the 
particle densities of ${\cal S}$ submitted to $U_{\alpha}^{{\rm B}}(\br)$ are indeed homogeneous and they 
are merely related to $c_{\rm B}$ \textsl{via} the neutrality condition~(\ref{neutralitybis}) 
valid for ${\cal S}^\ast$. Moreover, all particle correlations of both systems are identical.

\bigskip

If we insert expression~(\ref{GCPFidentity}) of $\Xi$ into definition~(\ref{FE})
of the free energy of ${\cal S}$ submitted to $U_{\alpha}^{{\rm B}}(\br)$, we obtain in the TL
\be
\la{FEnn}
{\cal F}\{\rho_1 , \rho_2 \} -
\frac{\beta}{2} \int_{\Lambda^2} \dd \br \dd \br' \frac{( q_1\rho_1 + q_2 \rho_2)^2}{|\br'-\br|} 
\sim {\cal F}^\ast \{\rho_1 , \rho_2 \}
\end{equation} 
while $c_{\rm B}$ has also been replaced by $-(q_1\rho_1 + q_2 \rho_2)$ thanks to the neutrality 
condition~(\ref{neutralitybis}). Thus, if we define, in general, 
the reduced free energy 
${\cal F}_{\rm red}$ of ${\cal S}$ by subtracting to ${\cal F}$ the self-electrostatic energy of the 
charge distribution $( q_1\rho_1(\br) + q_2\rho_2(\br))$, namely
\be
\la{FEreduced}
{\cal F}_{\rm red} = {\cal F}
- \frac{\beta}{2} \int_{\Lambda^2} \dd \br \dd \br' \frac{(q_1\rho_1(\br) + q_2\rho_2(\br))
( q_1\rho_1(\br') + q_2\rho_2(\br'))}{|\br'-\br|} \; ,
\end{equation} 
we find the remarkable identity
\be
\la{FEreducedbis}
{\cal F}_{\rm red}\{\rho_1 , \rho_2  \}  \sim {\cal F}^\ast \{\rho_1 , \rho_2 \} \; ,
\end{equation} 
which can be rewritten for the corresponding free-energy densities as 
\be
\la{FEreduceddensity}
f_{\rm red}(\rho_1,\rho_2, \beta) = f^\ast(\rho_1,\rho_2, \beta) \; .
\end{equation} 
Notice that this subtraction from the free-energy functional of the self-electrostatic energy 
was first introduced by Hohenberg and Kohn~\cite{HohKoh} for studying 
the quantum electron gas.

\subsection{Density functional expansions for almost homogeneous systems}

For states of ${\cal S}$ with slow spatial variations of the particle densities, the corresponding reduced
free-energy ${\cal F}_{\rm red} \{\rho_1(\cdot) , \rho_2(\cdot) \} $ can be expanded in powers 
of the gradients of $\rho_1(\br)$ and $\rho_2(\br)$. 
The leading term in that systematic expansion is purely local and reduces to 
\be
\la{FElocal}
\int \dd \br \; 
f^\ast(\rho_1(\br),\rho_2(\br), \beta)  \; ,
\end{equation} 
where we have used identity~(\ref{FEreduceddensity}) for the reduced free-energy density of an homogeneous system. 
The first correction, the so-called square-gradient term, reads~\cite{Yang1976,Bongiorno1976}
\be
\la{FEsqg}
\frac{1}{12} \sum_{\alpha,\gamma}\int \dd \br \; 
M_{\alpha\gamma}^\ast(\rho_1(\br),\rho_2(\br), \beta) 
\nabla \rho_{\alpha}(\br) \cdot \nabla \rho_{\gamma}(\br) \; ,
\end{equation} 
where $M_{\alpha\gamma}^\ast(\rho_1,\rho_2, \beta)$
is the second moment of the short-range part of the direct two-body correlations, 
namely $c_{\alpha\gamma}^{{\rm SR}}(r)=c_{\alpha\gamma}(r) + \beta q_{\alpha} q_{\gamma}/r$, 
for ${\cal S}^\ast$  with homogeneous densities $(\rho_1,\rho_2)$. Similarly to the 
emergence of the free-energy density of ${\cal S}^\ast$ in the purely local term~(\ref{FElocal}), 
the direct correlations of ${\cal S}^\ast$ arise in the square-gradient term because all 
the respective distribution functions 
of ${\cal S}^\ast$ and ${\cal S}$ with the same homogeneous densities 
are identical as established above.
Notice that for systems with short-range interactions, the square-gradient expansion of the 
free energy ${\cal F}$ involves second moments of the direct correlations themselves. Here, thanks to the 
subtraction~(\ref{FEreduced}) of the electrostatic self-energy, the square-gradient expansion of the 
reduced free energy ${\cal F}_{\rm red}$ involves second moments of the short-range part 
$c_{\alpha\gamma}^{{\rm SR}}(r)$, which do converge thanks to the large-distance 
behaviour $c_{\alpha\gamma}(r) \sim - \beta q_{\alpha} q_{\gamma}/r$ when $r \to \infty$.  

\bigskip

The second correction to the purely local contribution~(\ref{FElocal}) involves fourth-order spatial 
derivatives of the densities, and it has been explicitly computed in Ref.~\cite{Ala1983}. The corresponding 
local ingredients are fourth moments of two-, three- and four-body direct correlations of ${\cal S}^\ast$ with 
homogeneous densities. Higher-order corrections would exhibit similar structures with well-behaved local
ingredients defined for the same system.

\section{Linear response to a weak slowly-varying external charge distribution}
\la{LR}

We start with ${\cal S}$ in the absence of any applied external potential, 
namely that realization of ${\cal S}$ is nothing but the homogeneous neutral TCP.   
In a second step, let us introduce an external charge distribution
$c_{\rm ext}(\br)=q_{\rm ext} \exp(i \bk \cdot \br)$, with an infinitesimal amplitude 
$q_{\rm ext}$. Our aim here is to determine the induced charge density in ${\cal S}$ by DFT 
and compare to its linear response expression. This will provide  the required 
fourth moment sum rule for the charge correlations of the  homogeneous neutral TCP. 
In a first step, within DFT, we compute the density responses $\delta \rho_\alpha (\br)$ 
to the external potentials, $U_\alpha(\br)=  q_\alpha \varphi_{\rm ext}(\br)$ 
with $\varphi_{\rm ext}(\br)$ the electrostatic potential created by $c_{\rm ext}(\br)$, at 
leading order in $q_{\rm ext}$ and in the limit of small wave-numbers $k \to 0$.

\subsection{Analysis within density functional theory}

Since the applied external potential varies on an infinitely large scale length, the particle 
densities should also display infinitely slow spatial variations. Then, the free-energy functional 
can be replaced by its density-gradient expansion introduced above, namely
\begin{multline}
\la{FEdge}
{\cal F}\{\rho_1(\cdot) , \rho_2(\cdot) \} = \frac{\beta}{2} 
\int \dd \br \dd \br' \frac{(q_1\rho_1(\br) + q_2\rho_2(\br))
( q_1\rho_1(\br') + q_2\rho_2(\br'))}{|\br'-\br|} \\ 
+ \int \dd \br \; 
f^\ast(\rho_1(\br),\rho_2(\br), \beta) 
+ \frac{1}{12} \sum_{\alpha,\gamma}\int \dd \br \; 
M_{\alpha\gamma}^\ast(\rho_1(\br),\rho_2(\br), \beta) 
\nabla \rho_{\alpha}(\br) \cdot \nabla \rho_{\gamma}(\br) + ... \; ,
\end{multline}   
where the terms left over do not contribute to the deviations 
$\delta \rho_\alpha (\br)$ at the considered lowest orders in $k$, as 
shown further on. The fundamental equation~(\ref{functionalderivativeFE}) of DFT then becomes for each species,
\begin{multline}
\la{FunctionalEP}
\frac{ \partial f^\ast}{\partial \rho_1} 
-\frac{1}{6} \left[ M_{11}^\ast \Delta \rho_1(\br)
+ M_{12}^\ast \Delta \rho_2(\br)\right] \\
-\frac{1}{12} \left[ \frac{ \partial  M_{11}^\ast }{\partial \rho_1} (\nabla \rho_1(\br))^2 
+ 2 \frac{\partial  M_{11}^\ast }{\partial \rho_2} \nabla \rho_1(\br) \cdot \nabla \rho_2(\br) 
+(2 \frac{\partial  M_{12}^\ast }{\partial \rho_2} - \frac{\partial  M_{22}^\ast }{\partial \rho_1}) (\nabla \rho_2(\br))^2 \right] 
+... \\
= \beta \mu_1  -\beta  q_1 \varphi_{\rm tot}(\br)
\end{multline}
and 
\begin{multline}
\la{FunctionalEM}
\frac{ \partial f^\ast}{\partial \rho_2} 
-\frac{1}{6} \left[ M_{22}^\ast \Delta \rho_2(\br)
+ M_{12}^\ast \Delta \rho_1(\br)\right] \\
-\frac{1}{12} \left[ \frac{ \partial  M_{22}^\ast }{\partial \rho_2} (\nabla \rho_2(\br))^2 
+ 2 \frac{\partial  M_{22}^\ast }{\partial \rho_1} \nabla \rho_1(\br) \cdot \nabla \rho_2(\br) 
+(2 \frac{\partial  M_{12}^\ast }{\partial \rho_1} - \frac{\partial  M_{11}^\ast }{\partial \rho_2}) (\nabla \rho_1(\br))^2 \right] 
+... \\
= \beta \mu_2  -\beta  q_2 \varphi_{\rm tot}(\br) \; ,
\end{multline}
where $\varphi_{\rm tot}(\br)$ is the total electrostatic potential created by the charge distribution 
$(q_1\rho_1(\br) + q_2\rho_2(\br) + c_{\rm ext}(\br))$.
In Eqts.~(\ref{FunctionalEP},\ref{FunctionalEM}) all involved quantities of ${\cal S}^\ast$ are evaluated for homogeneous densities identical to the local densities 
$(\rho_1(\br),\rho_2(\br))$ and inverse temperature $\beta$. Moreover, all terms left over  
involve at least fourth-order spatial derivatives of $\rho_1(\br)$ and $\rho_2(\br)$.

In order to compute the induced densities at lowest order in $q_{\rm ext}$, we can linearize  
Eqts.~(\ref{FunctionalEP},\ref{FunctionalEM}) with respect to $\delta \rho_\alpha (\br)$. The third terms in the 
l.h.s. of those Eqts. do not contribute anymore since they are at least of order $q_{\rm ext}^2$. The resulting 
deviations take the form of plane waves, like the forcing external charge $c_{\rm ext}(\br)$, namely
\be
\la{deviation}
\delta \rho_1 (\br) = A_1(k) \exp(i \bk \cdot \br) \;\;\; \text{and} \;\;\; \delta \rho_2 (\br) = A_2(k) \exp(i \bk \cdot \br)
\end{equation}
where the amplitudes $A_\alpha(k)$ are proportional to $q_{\rm ext}$. The total electrostatic potential
$\varphi_{\rm tot}(\br)$ satisfies Poisson equation
\be
\la{Poisson}
\Delta \varphi_{\rm tot}(\br) = -4 \pi \left[ q_1\rho_1(\br) + q_2\rho_2(\br) + c_{\rm ext}(\br) \right]
=-4 \pi \left[ q_1 \delta\rho_1(\br) + q_2 \delta \rho_2(\br) + c_{\rm ext}(\br) \right]
\end{equation}
where the second equality follows from the overall neutrality of the unperturbed system ${\cal S}$ . In order to eliminate 
$\varphi_{\rm tot}(\br)$ in favor of the induced density deviations, it is then sufficient to
take the Laplacian of the linearized versions of 
Eqts.~(\ref{FunctionalEP},\ref{FunctionalEM}). This provides  
\begin{multline}
\la{deviationplus}
\left[4\pi\beta q_1^2 + \chi_{11}^{-1}k^2 + a_{11} k^4 + O(k^6)\right] A_1(k) + 
\left[4\pi\beta q_1q_2 + \chi_{12}^{-1}k^2 + a_{12} k^4 + O(k^6)\right]A_2(k) \\ 
= -4\pi\beta q_1 \delta q_{\rm ext}
\end{multline}
\begin{multline}
\la{deviationminus}
\left[4\pi\beta q_1q_2 + \chi_{21}^{-1}k^2 +  a_{21} k^4 + O(k^6)\right] A_1(k)+
\left[4\pi\beta q_2^2 + \chi_{22}^{-1}k^2 +  a_{22} k^4 + O(k^6)\right]A_2(k) \\
 = -4\pi\beta q_2 \delta q_{\rm ext}
\end{multline}
with $\chi_{\alpha\gamma}^{-1}= \partial^2 f^\ast/\partial \rho_\alpha \partial \rho_\gamma$ and  
$a_{\alpha\gamma}=M_{\alpha\gamma}^\ast/6$. Those reference quantities are evaluated for the set 
$(\rho_1,\rho_2)$ ensuring overall neutrality of the unperturbed system ${\cal S}$. Notice that if the 
thermodynamic function $\chi_{\alpha\gamma}^{-1}$ is specific to the enlarged auxiliary system ${\cal S}^\ast$, 
the microscopic second moments $a_{\alpha\gamma}$ entirely depend on the direct correlations of the genuine 
system ${\cal S}$ of interest.

The linear Eqts.~(\ref{deviationplus},\ref{deviationminus}) are straightforwardly solved in terms
of the determinant of the associated two by two matrix which reads 
\begin{multline}
\la{determinant}
D(k)= 4\pi\beta (q_2^2 \chi_{11}^{-1} + q_1^2 \chi_{22}^{-1} - 2 q_1q_2 \chi_{12}^{-1}) k^2 \\
+ \left[\chi_{11}^{-1}\chi_{22}^{-1} - \chi_{12}^{-2} + 4\pi\beta (q_2^2 a_{11} + q_1^2 a_{22} - 2 q_1q_2 a_{12}) \right] k^4 
+ O(k^6) \;  . 
\end{multline}
The amplitudes $A_\alpha(k)$ are then found to be
\be
\la{amplitudeplus}
A_1(k)=\frac{4\pi\beta}{D(k)}\left[ (q_2 \chi_{12}^{-1} - q_1 \chi_{22}^{-1}) k^2 +(q_2 a_{12} - q_1 a_{22})k^4 + O(k^6)\right]
\delta q_{\rm ext}
\end{equation}
and 
\be
\la{amplitudeminus}
A_2(k)=\frac{4\pi\beta}{D(k)}\left[ (q_1 \chi_{12}^{-1} - q_2 \chi_{11}^{-1}) k^2 +(q_1 a_{12} - q_2 a_{11})k^4 + O(k^6)\right]
\delta q_{\rm ext} \; .
\end{equation}
Therefore the proportionality coefficient between a given amplitude and $\delta q_{\rm ext}$, behaves in the limit $k \to 0$ as 
a constant, which depends only on the thermodynamic quantities $\chi_{\alpha\gamma}^{-1}$, plus a term of order $k^2$ which 
depends on both $\chi_{\alpha\gamma}^{-1}$ and $a_{\alpha\gamma}$. Now, if we form the induced charge density 
\be
\la{inducedcharge}
\delta c (\br)= q_1 \delta \rho_1 (\br) + q_2 \delta \rho_2 (\br) = c_{\rm ind}(k) \exp(i \bk \cdot \br)
\end{equation}
with the charge amplitude
\be
\la{chargeamplitude}
c_{\rm ind}(k) = q_1 A_1(k) + q_2 A_2(k) \; ,
\end{equation}
we find 
\be
\la{chargeamplitudebis}
c_{\rm ind}(k) = -\delta q_{\rm ext} 
[1-\frac{(\chi_{11}^{-1}\chi_{22}^{-1}-\chi_{12}^{-2})}
{4\pi\beta (q_2^2 \chi_{11}^{-1}+q_1^2 \chi_{22}^{-1}-2 q_1q_2 \chi_{12}^{-1})}\; k^2 + O(k^4)] \; .
\end{equation}
Remarkably, the proportionality coefficient between the induced and external charges goes to $-1$ when 
$k \to 0$, in relation with perfect screening properties, as discussed further. Furthermore the term 
of order $k^2$ in its small-$k$ expansion now depends only on the thermodynamical functions $\chi_{\alpha\gamma}^{-1}$, 
and no longer on the microscopic quantities $a_{\alpha\gamma}$.

\subsection{The fourth moment sum rule} 

The resulting induced charge density, can be also determined 
within linear response theory, which provides
\be
\la{LinearResponse}
\delta c (\br) =  -\frac{4 \pi \beta}{k^2 } \tilde{S}(k) 
\delta q_{\rm ext} \exp(i \bk \cdot \br)
\end{equation}
In the linear response formula~(\ref{LinearResponse}), $\tilde{S}(k)$ is the Fourier 
transform of the charge correlations of the unperturbed 
system ${\cal S}$, i.e. the homogeneous neutral TCP, 
\be
\la{CC}
\tilde{S}(k) = \int \dd \br  \exp(i \bk \cdot \br) 
\left[\sum_{\alpha,\gamma} q_\alpha q_\gamma \rho_{\alpha\gamma}(r)+\sum_\alpha q_\alpha^2 \rho_\alpha \delta(\br) \right]
\end{equation}
with $\rho_{\alpha\gamma}(r)$ the two-body probability density for the spatial configuration 
where one particle of species $\alpha$ is fixed at the origin, while another particle of species 
$\gamma$ is fixed at $\br$.  

The small-$k$ expansion of the amplitude $c_{\rm ind}(k)$ can be inferred from 
the linear response formula~(\ref{LinearResponse}) by inserting the corresponding expansion of 
$\tilde{S}(k)$, 
\be
\la{CCexpansion}
\tilde{S}(k)=I_0 + I_2 k^2 + I_4 k^4 + ...\; ,
\end{equation}
which only involves powers of $k^2$ thanks to the expected exponential decay of charge correlations in 
real space. If we compare the resulting expansion of $c_{\rm ind}(k)$ with the DFT result~(\ref{chargeamplitudebis}), 
we readily find 
\be
\la{SR02}
I_0=0 \;\;\; \text{and} \;\;\; I_2=\frac{1}{4\pi\beta}
\end{equation}
which follow from respectively the absence of a $1/k^2$-term, and the identification of the constant terms. The vanishing of 
$I_0$ accounts for the perfect screening of internal charges. The universal value of $I_2$, first 
demonstrated a long ago by Stillinger and Lovett~\cite{StiLov}, ensures the perfect screening of weak external charges. 
Beyond those well-known results for the zeroth and second moments of $S(r)$, the DFT expression~(\ref{chargeamplitudebis}) 
also provides a new sum rule for the fourth moment, namely 
\be
\la{SR4}
I_4 = -\frac{\rho^2 }{(4\pi (q_1-q_2))^2 \beta^3} (\chi_{11}^{-1}\chi_{22}^{-1}-\chi_{12}^{-2}) \chi_T \; ,
\end{equation}
which follows from the identification of the $k^2$-terms. The compressibility $\chi_T$ emerges in that sum rule, thanks to the 
identity~(\ref{IsoCompbis}) rewritten in terms of the charges $q_1=Z_1 q$ and $q_2=-Z_2 q$.

\subsection{Related sum rules for other models}  

Let us first consider the case of the OCP. A fourth moment sum rule for the 
corresponding charge correlations $S_{\rm OCP}(r)$ was derived by Vieillefosse and Hansen~\cite{Vieille} 
through a macroscopic analysis of fluctuations. In their textbook~\cite{HansenMacDo}, Hansen and Mac Donald 
propose a simple derivation which is similar to ours. They compute the charge density induced by a weak 
external plane wave charge distribution within an hydrostatic approach, where 
the force associated with the local pressure gradient is balanced by the total electrostatic 
force created by both the external and induced charges. Notice that the corresponding equation 
can be merely obtained by taking the gradient of the fundamental DFT equation~(\ref{FunctionalEP}) 
restricted to a single species and where all non-local contributions, including that involving the second moment of the direct correlations, are omitted. Moreover, the corresponding $f^\ast$ can then be obviously replaced by $f_{\rm OCP}$. 
The fourth moment of $S_{\rm OCP}(r)$ then reduces to~\cite{Vieille,HansenMacDo}, 
\be
\la{SR4OCP}
I_4^{\rm OCP} = -\frac{1 }{(4\pi q \rho)^2 \beta \chi_T^{\rm OCP}} \; .
\end{equation}
Notice that this expression has been recovered through manipulations of the BGY hierarchy, for pure Coulomb 
interactions~\cite{Vieille1985} and also including short-range interactions~\cite{BraVieille}.

The OCP result has been extended to a multicomponent ionic mixture (MIM) of all positive point charges immersed 
in a rigid neutralizing background~\cite{Suttorp1987,Suttorp2008}. Interestingly, the derivation is intrinsic 
and does not rely on the response of the system to a weak external charge distribution. Like the analysis~\cite{Vieille1985}
carried out for the OCP, it is based on suitable 
manipulations of the BGY hierarchy equations for the distribution functions of the infinite homogeneous neutral system. 
\textit{A priori} the derivation is only valid for pure Coulomb interactions, without any short range regularization
which is unnecessary here since all mobile charges repel together. It makes an explicit use of the 
remarkable homogeneity property of the resulting pair interactions. 
The fourth moment of charge correlations in real space is then given by formula (7.3) in Ref.~\cite{Suttorp2008}, 
which reduces in three dimensions ($d=3$) to
\begin{equation}
\la{Suttorp}
\int \dd \br \; r^4 \; S_{\rm MIM}(r) = -\frac{120}
{\beta \sum_{\alpha,\gamma} q_{\alpha} q_{\gamma} \partial \rho_\gamma/\partial \mu_\alpha}
\end{equation}
where we have used that $q_{v}=\sum_\alpha q_{\alpha} \rho_\alpha $, while 
$\mu_\alpha = \beta^{-1}\partial f_{\rm MIM}/\rho_\alpha$. Each partial derivative $\partial \rho_\gamma/\partial \mu_\alpha$ 
is computed by fixing the inverse temperature $\beta$ as well as all $\mu_{\delta}$'s with $\delta \neq \alpha$. 
Straightforward manipulations of the multi-variable functions $\rho_\gamma(\beta,\{\mu_\alpha \})$ and 
$\mu_\alpha(\beta,\{\rho_\gamma \})$ allow us to express all partial derivatives
$\partial \rho_\gamma/\partial \mu_\alpha$ in terms of partial derivatives 
$\partial \mu_\alpha/\partial \rho_\gamma$. 
In the binary case, we find
\begin{multline}
\la{DCinversion}
\frac{\partial \rho_1}{\partial \mu_1} = \frac{\partial \mu_2}{\partial \rho_2} 
\left[\frac{\partial \mu_1}{\partial \rho_1}\frac{\partial \mu_2}{\partial \rho_2}
-\frac{\partial \mu_1}{\partial \rho_2}\frac{\partial \mu_2}{\partial \rho_1}\right]^{-1} \;\; 
; \;\;
\frac{\partial \rho_2}{\partial \mu_2} = \frac{\partial \mu_1}{\partial \rho_1} 
\left[\frac{\partial \mu_1}{\partial \rho_1}\frac{\partial \mu_2}{\partial \rho_2}
-\frac{\partial \mu_1}{\partial \rho_2}\frac{\partial \mu_2}{\partial \rho_1}\right]^{-1} \; \\
\frac{\partial \rho_1}{\partial \mu_2} = -\frac{\partial \mu_1}{\partial \rho_2} 
\left[\frac{\partial \mu_1}{\partial \rho_1}\frac{\partial \mu_2}{\partial \rho_2}
-\frac{\partial \mu_1}{\partial \rho_2}\frac{\partial \mu_2}{\partial \rho_1}\right]^{-1} \;\; 
; \;\;  \frac{\partial \rho_2}{\partial \mu_1} = -\frac{\partial \mu_2}{\partial \rho_1} 
\left[\frac{\partial \mu_1}{\partial \rho_1}\frac{\partial \mu_2}{\partial \rho_2}
-\frac{\partial \mu_1}{\partial \rho_2}\frac{\partial \mu_2}{\partial \rho_1}\right]^{-1} \; .
\end{multline}
Using the identity
\begin{equation}
\la{RelationI4}
I_4=\frac{1}{120} \int \dd \br \; r^4 \; S(r)  
\end{equation}
and inserting relations~(\ref{DCinversion}) into formula~(\ref{Suttorp}), we find that the corresponding 
$I_4^{\rm BIM}$ exactly coincides with our general expression~(\ref{SR4}) 
specified to the BIM, where the free-energy density $f^\ast$ 
merely reduces to $f_{\rm BIM}$. Indeed, our derivation also applies to the BIM where $q_1$ and 
$q_2$ now have the same sign, while the auxiliary system ${\cal S}^\ast$ becomes identical to the genuine 
BIM of interest with the background charge density $c_{\rm B}=-q_{v}=-\sum_\alpha q_{\alpha} \rho_\alpha$.   

\subsection{About other approaches} 

To our knowledge, in 
the literature, there exist two attempts to derive a sum rule for the fourth moment of the charge correlations 
of the TCP. First, the 
hydrodynamic approach carried out in Ref.~\cite{Gia} provides an expression for the fourth moment, 
different from formula~(\ref{SR4}), which involves 
ill-defined thermodynamic quantities as well as particle masses. Its validity is then quite doubtful, in particular 
because classical equilibrium charge correlations do not depend on particle masses. 

Second, van Beijeren and Felderhof~\cite{vBF1979} proceed to an intrinsic analysis of charge correlations
within the grand-canonical ensemble, where they combine the Ornstein-Zernicke equations with DFT manipulations. 
In agreement with results previously derived by Mitchell \textit{et al.}~\cite{Mit1977}, who shown that 
the fourth moment cannot be expressed in terms of thermodynamic 
quantities of the sole TCP, they find that it is necessary to introduce non-neutral states
of the TCP which can be realized through the application of a suitable external potential. However, they 
did not provide any scheme which determines that external potential. Thus their free-energy density $f^0$, from which the thermodynamical chemical potentials are inferred through the usual identity written in formula (3.12) of Ref.~\cite{vBF1979}, 
remains a formal quantity, with no prescriptions for explicit calculations. This 
ambiguity might explain why their work is not always cited. According to our analysis, it can 
be easily clarified as follows. In fact, as shown in Section~\ref{SI}, the 
external potential mentioned in Ref.~\cite{vBF1979} is nothing but our 
potential $U_\alpha^{{\rm B}}(\br)= q_\alpha \varphi_{\rm B}(\br)$ 
where $\varphi_{\rm B}(\br)$ is the electrostatic potential created by an homogeneous background density. 
Therefore, $f^0$ is identical to our free-energy density $f^\ast$ of the TCP immersed in an uniform 
rigid background. Then, the relation between partial derivatives~(\ref{DCinversion}) 
allows us to exactly recast formula~(6.26) of Ref.\cite{vBF1979} as our expression~(\ref{SR4}), 
similarly to what occurs for the corresponding formula obtained for the BIM by Suttorp~\cite{Suttorp2008}.

\section{Asymptotic expansions at low densities} 
\la{LDE}

It is instructive to check the fourth moment sum rule for specific models and 
various ranges of thermodynamical parameters. Here, we consider the 
model of charged soft spheres with the pair interaction $u_{\alpha\gamma}(r)$ given by 
formula~(\ref{soft}). First, we briefly describe how the pair correlations 
of ${\cal S}^\ast$ can be represented by an infinite series of resummed Mayer 
graphs. Such resummed diagrammatics constitute a quite suitable framework for 
deriving low-density expansions of the quantities of interest. From the 
diagrammatic representation of charge correlations, we infer the low-density expansion 
of $I_4$ defined as the coefficient of the $k^4$-term in the small-$k$ expansion~(\ref{CCexpansion}) 
of $\tilde{S}(k)$. The diagrammatics for the pair correlations also 
give access to the free energy density $f^\ast$ through thermodynamical identities. 
The low-density expansion of the thermodynamical expression~(\ref{SR4}) is then 
computed, and it is shown to exactly match that of $I_4$, as expected.

\subsection{Exploiting the principle of topological reduction}

Let  $\rho_{\alpha\gamma,\rm T}^\ast(r) =  \rho_{\alpha\gamma}^\ast(r) - \rho_\alpha \rho_\gamma$
be the truncated pair distribution functions of ${\cal S}^\ast$, also called pair correlations, for 
an arbitrary set of densities $(\rho_1,\rho_2)$. 
As argued above, the distributions function of ${\cal S}^\ast$, which includes 
a background with charge density $c_{\rm B}= -(q_1 \rho_1 + q_2 \rho_2)$, are identical to that of 
of a purely two-component system where the mobile particles 
are submitted to the external potential~(\ref{ExtPotUniform}) $U_{\alpha}^{{\rm B}}(\br)$ created by
the background. Therefore, pair correlations $\rho_{\alpha\gamma,\rm T}^\ast(r)$ 
are represented by series of Mayer diagrams~\cite{Mayer} made with two root (white) points respectively fixed 
at the origin $\textbf{0}$ and at $\br$, and an arbitrary number of black points whose positions are integrated over. 
Each point carries a statistical weight 
\be
\la{fugacityalphaB}
z_\alpha  = \frac{\exp [\beta (\mu_\alpha -U_{\alpha}^{{\rm B}}) ]}
{(2 \pi \lambda_\alpha^2)^{3/2}} \;,  
\end{equation} 
while two points are connected by at most one Mayer bond   
\be
\la{MayerBond}
b_{\rm M}=\exp (-\beta u_{\alpha\gamma}) -1 \; .
\end{equation}
Each diagram is simply connected, namely there exists at least one path connecting two arbitrary points. 

\bigskip

The previous Mayer diagrams are difficult to handle because the fugacity weights~(\ref{fugacityalphaB}) are inhomogeneous and 
depend on the positions of the points. A great simplification can be achieved by virtue of the principle 
of topological reduction, nicely exposed in Ref.~\cite{HansenMacDo}, which consists in removing all articulation points. 
An articulation point is such that there exists at least one subdiagram attached to it and not connected to 
the rest of the diagram. In other words, the suppression of the articulation point leaves that 
subdiagram disconnected from the two root points. If one sums all those subdiagrams 
attached to a given articulation point, all articulation points are removed, while simultaneously  
all fugacity weights~(\ref{fugacityalphaB}) are replaced by density weights  $\rho_\alpha$~\cite{HansenMacDo}. Furthermore, 
the topological structure of the diagrams is conserved through that reduction. Accordingly, the 
pair correlations $\rho_{\alpha\gamma,\rm T}^\ast(r)$ are represented by Mayer diagrams made with 
the two root points fixed at $\textbf{0}$ and $\br$, and an arbitrary number of black points, where the 
point statistical weights are now the densities $\rho_\alpha$. Two point are still connected at most 
by one Mayer bond~(\ref{MayerBond}). Each diagram is again simply connected but is now free of any articulation point.    

\bigskip

Thanks to the translational invariance of both density weights and Mayer bonds, the Mayer density diagrams 
reveal quite useful for explicit calculations as described further. Notice that, remarkably, the background 
does not show in such diagrams, its effects being implicitly and entirely taken into account by the introduction of 
the homogeneous densities $\rho_\alpha$
 
\subsection{Abe-Meeron resummations}

Because of the long-range non-integrable decay of two-body interactions $u_{\alpha\gamma}$, every Mayer diagram 
diverges. All those divergencies can be removed via chain resummations, as first noticed by Mayer~\cite{Mayer1950} and 
Salpeter~\cite{Salpeter}, and then performed in a systematic way for the whole diagrammatical series
by Abe~\cite{Abe} and Meeron~\cite{Meeron}. A simplified presentation of that method 
can be found in Refs.~\cite{ACP1994} and \cite{ABCM2003}. It starts with the decomposition  of each Mayer 
bond~(\ref{MayerBond}) as 
\be
\la{BondDec}
b_{\rm M} = b_{\rm M}^{\rm T}-\beta q_\alpha q_\gamma v_{\rm C} 
\end{equation}
with the truncated bond
\be
\la{BondT}
b_{\rm M}^{\rm T}= \exp (-\beta u_{\alpha\gamma}) -1 + \beta q_\alpha q_\gamma v_{\rm C} 
\end{equation} 
and the Coulomb potential $v_{\rm C}(r)=1/r$. After inserting the decomposition~(\ref{BondDec}) 
into every Mayer diagram, one proceeds to systematic resummations of convolution chains of  
Coulomb bonds $-\beta q_\alpha q_\gamma v_{\rm C}$. Thanks to remarkable combinatorial properties~\cite{ABCM2003}, 
all those resummations can be performed in terms of a single effective potential, which is nothing but 
the well-known Debye potential
\be
\la{Debye}
\phi_{\rm D} (r)= \frac{\exp(-\kappa_{\rm D}r)}{r}
\end{equation}  
with the Debye inverse length $\kappa_{\rm D}=(\sum_\alpha 4 \pi \beta q_\alpha^2 \rho_\alpha)^{1/2}$. The chain resummations 
give raise to two bonds~\cite{ACP1994}, the Debye bond
\be
\la{BondDebye}
b_{\rm D} = -\beta q_\alpha q_\gamma \phi_{\rm D} 
\end{equation}  
and the short-range dressed bond 
\be
\la{BondSRD}
b_{\rm R}= \exp (-\beta (u_{\alpha\gamma}^{\rm SR} + q_\alpha q_\gamma \phi_{\rm D})) -1 + \beta q_\alpha q_\gamma \phi_{\rm D} \, , 
\end{equation} 
with the short-range part of pair interactions  
$u_{\alpha\gamma}^{\rm SR}=u_{\alpha\gamma} - q_\alpha q_\gamma v_{\rm C}$. The topological structure of the genuine 
Mayer diagrams remain unchanged, with bonds which can be either $b_{\rm D}$ or $b_{\rm R}$, and with the 
additional rule excluding convolutions $b_{\rm D} \ast b_{\rm D}$ in order to avoid double counting.

Within the Abe-Meeron resummations, the genuine whole set of Mayer diagrams representing $\rho_{\alpha\gamma,\rm T}^\ast(r)$ 
is then exactly transformed into 
\be
\la{MayerResummed}
\rho_{\alpha\gamma,\rm T}^\ast(r) = \rho_\alpha \rho_\gamma \sum_{\cal G} \frac{1}{S_{\cal G}} 
\int \left[ \prod_{i=1}^n \sum_{\alpha_i} \dd \br_i \rho_{\alpha_i} \right] 
\left[ \prod{b_{\rm D}} \prod{b_{\rm R}} \right]_{\cal G} \; .
\end{equation}
The so-called prototype graphs ${\cal G}$ are made with 
the two root points respectively fixed at $\textbf{0}$ and $\br$, and an arbitrary number of $n$ black points
with density weights. Two point are connected at most 
by one bond~(\ref{BondDebye}) or (\ref{BondSRD}). Each diagram is simply connected, with no articulation points, while
convolutions $b_{\rm D} \ast b_{\rm D}$ are forbidden. The symmetry factor $S_{\cal G}$ is defined as the number 
of permutations of labelled black points which leave the product of bonds 
$\left[ \prod{b_{\rm D}} \prod{b_{\rm R}} \right]_{\cal G}$ unchanged. 
The summation is carried out over all topologically different graphs ${\cal G}$, including the simplest 
graph with no black points. 

In the diagrammatic representation~(\ref{MayerResummed}), the contribution of every graph ${\cal G}$ is finite. 
Indeed, at large distances, 
integrability is ensured by the fast decays of both the Debye potential and the short-range part of pair interactions. 
At short distances, the Debye bond remains integrable despite its $1/r$ singularity, while the short-range dressed 
bond includes the short-range regularization which also guarantees its integrability. We stress that 
representation~(\ref{MayerResummed}) holds for any set $(\rho_1,\rho_2)$ of densities, and then appears to be 
quite useful for computing equilibrium quantities of ${\cal S}^\ast$. Moreover it is valid for any short-range regularization 
$u_{\alpha\gamma}^{\rm SR}$, including of course that describing soft or hard spheres.      

\subsection{Charge correlations}

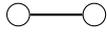
\begin{figure}
 \begin{center}
	\begin{tikzpicture}
		\node[blanc] (R0) at (0,0) {};
		\node[blanc] (R1) at (1,0) {};
		\drawLinkFC{R0}{R1};
	\end{tikzpicture}
 \end{center}
\caption{The Debye diagram in the resummed diagrammatic representation~(\ref{MayerResummed}) of particle correlations. The 
two root (white points) are fixed at $\textbf{0}$ and $\br$ respectively. The straight line represents a 
Debye bond $b_{\rm D}$~(\ref{BondDebye}). }
\label{DebyeDiagram}
\end{figure}

The Fourier transform~(\ref{CC}) of the charge correlations of the homogeneous TCP can be recast as 
\be
\la{CCbis}
\tilde{S}(k) = \sum_{\alpha,\gamma} q_\alpha q_\gamma \tilde{\rho}_{\alpha\gamma}(k)+\sum_\alpha q_\alpha^2 \rho_\alpha \;
\end{equation}
where the Fourier transform $\tilde{\rho}_{\alpha\gamma}(k)$ of pair correlations is given by the sum of the Fourier transforms 
of the contributions of all graphs ${\cal G}$ in the representation~(\ref{MayerResummed}). Let us first consider the contribution of 
the simplest graph ${\cal G}_{\rm D}$ shown in Fig.~\ref{DebyeDiagram}, where the two root points are connected by a Debye bond. 
Its contribution to  $\tilde{S}(k)$ 
added to the constant term $\sum_\alpha q_\alpha^2 \rho_\alpha$ in the formula~(\ref{CCbis}) provides the well-known 
Debye charge correlations
\be
\la{DebyeCC}
\tilde{S}_{\rm D}(k) = \frac{\kappa_{\rm D}^2}{4 \pi \beta} \frac{k^2}{k^2+\kappa_{\rm D}^2} \; ,
\end{equation}
which can be derived within a mean-field treatment of correlations, without any diagrammatic considerations. Now we stress 
that $\tilde{S}_{\rm D}(k) $ saturates the first two moments sum rules for $I_0$ and $I_2$, since 
$\tilde{S}_{\rm D}(k) \sim k^2/(4 \pi \beta)$ when $k \to 0$. Therefore all the remaining graphs in the 
representation~(\ref{MayerResummed}) give no contributions to $I_0$ and $I_2$. That remarkable property is related to the 
following reorganization of the series of graphs, which turns out to be also quite useful for computing the fourth 
moment $I_4$. 

\begin{figure}
	\begin{center}
		\begin{tikzpicture}
			\node[blanc] (R0) at (0,0) {};
			\node[noir] (R1) at (1,0) {};
			\drawLinkFC{R0}{R1};
			\node[blanc] (R2) at (2,0) {};
			\drawLinkFA{R1}{R2}{${\cal G}_{\rm d}$};
		\end{tikzpicture}
		\quad\quad
		\begin{tikzpicture}
			\node[blanc] (R0) at (0,0) {};
			\node[noir] (R1) at (1,0) {};
			\drawLinkFA{R0}{R1}{${\cal G}_{\rm d}$};
			\node[blanc] (R2) at (2,0) {};
			\drawLinkFC{R1}{R2};
		\end{tikzpicture}
		\quad\quad
		\begin{tikzpicture}
			\node[blanc] (R0) at (0,0) {};
			\node[noir] (R1) at (1,0) {};
			\drawLinkFC{R0}{R1};
			\node[noir] (R2) at (2,0) {};
			\drawLinkFA{R1}{R2}{${\cal G}_{\rm d}$};
			\node[blanc] (R3) at (3,0) {};
			\drawLinkFC{R2}{R3};
		\end{tikzpicture}
	\end{center}
	\caption{The three dressed Debye diagrams associated with a given diagram ${\cal G}_{\rm d}$ 
	in the resummed diagrammatic representation~(\ref{MayerResummed}) of particle correlations. } 
	\label{DressedDebyeDiagram}
\end{figure}
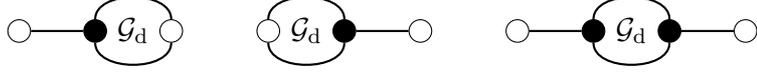

Let ${\cal G}_{\rm d}$ be a graph in the representation~(\ref{MayerResummed}) such that the root points $\textbf{0}$ 
and $\br$ are not connected to rest of the diagram by a single Debye bond $b_{\rm D}$, or in other words each 
root point is connected to the rest of the diagram by either a bond $b_{\rm R}$ or two bonds. Such a graph 
can be dressed by Debye bonds in the sense that the three graphs shown in Fig.~\ref{DressedDebyeDiagram} also intervene 
in the representation~(\ref{MayerResummed}). 
In ${\cal G}_{\rm Dd}$ (${\cal G}_{\rm dD}$), the black point $\br_1$ 
is connected to the root point $\textbf{0}$ ( $\br$)  by a Debye bond $b_{\rm D}$, while the 
subdiagram connecting that black point to the 
other root point $\br$ ( $\textbf{0}$ ) is identical to ${\cal G}_{\rm d}$ itself. In ${\cal G}_{\rm DdD}$, 
the two black points $\br_1$ and $\br_2$ are respectively connected to the root points $\textbf{0}$  and $\br$ 
by Debye bonds $b_{\rm D}$, while they are connected together by a subdiagram identical to ${\cal G}_{\rm d}$.
Clearly, all possible graphs ${\cal G}_{\rm d}$ together with their dressed Debye family generate all 
graphs in the representation~(\ref{MayerResummed}) beyond the Debye graph ${\cal G}_{\rm D}$.

Now, let us consider the total contribution to representation~(\ref{MayerResummed}) 
of a given graph ${\cal G}_{\rm d}$ and of its dressed Debye diagrams. 
After defining 
\be
\la{IntBlack}
K_{\alpha\gamma,{\cal G}_{\rm d}}(r) = \int \left[ \prod_{i=1}^n \sum_{\alpha_i} \dd \br_i \rho_{\alpha_i} \right] 
\left[ \prod{b_{\rm D}} \prod{b_{\rm R}} \right]_{{\cal G}_{\rm d}}
\end{equation}
and noticing that all four graphs ${\cal G}_{\rm d}$, ${\cal G}_{\rm Dd}$, ${\cal G}_{\rm dD}$ and ${\cal G}_{\rm DdD}$
have the same symmetry factor $S_{{\cal G}_{\rm d}}$, we can rewrite that total contribution as 
$\rho_\alpha\rho_\gamma/S_{{\cal G}_{\rm d}}$ times
\begin{multline}
\la{DressedFamily}
 K_{\alpha\gamma,{\cal G}_{\rm d}}(r)
- \sum_{\alpha_1} \rho_{\alpha_1} \int \dd \br_1 [\beta q_\alpha q_{\alpha_1} 
\phi_{\rm D}(r_1) K_{\alpha_1\gamma,{\cal G}_{\rm d}}(|\br -\br_1|) + 
K_{\alpha\alpha_1,{\cal G}_{\rm d}}(r_1) \beta q_{\alpha_1} q_\gamma 
\phi_{\rm D}(|\br -\br_1|)] \\
+ \sum_{\alpha_1,\alpha_2} \rho_{\alpha_1} \rho_{\alpha_2} \int \dd \br_1 \dd \br_2 \beta q_\alpha q_{\alpha_1} 
\phi_{\rm D}(r_1) K_{\alpha_1\alpha_2,{\cal G}_{\rm d}}(|\br_2 -\br_1|) \beta q_{\alpha_2} q_\gamma 
\phi_{\rm D}(|\br -\br_2|)
\end{multline}
The corresponding contribution to $\tilde{S}(k) $ in formula~(\ref{CCbis}) can be readily computed by using the convolution 
theorem and $\tilde{\phi}_{\rm D}(k)=4\pi/(k^2 + \kappa_{\rm D}^2)$, with the result
\be
\la{DressedFamilybis}
\frac{k^4}{(k^2 + \kappa_{\rm D}^2)^2} \sum_{\alpha,\gamma} \rho_\alpha\rho_\gamma q_{\alpha} q_\gamma
\tilde{K}_{\alpha\gamma,{\cal G}_{\rm d}}(k) 
\end{equation}
divided by the symmetry factor $S_{{\cal G}_{\rm d}}$. Since that expression is 
at least of order $k^4$ when $k \to 0$ for any ${\cal G}_{\rm d}$, all the graphs beyond the 
Debye graph ${\cal G}_{\rm D}$ do not contribute neither to $I_0$, nor to $I_2$. 
Moreover, because of the prefactor of order $k^4$, the resulting contribution to 
$I_4$ makes $\tilde{K}_{\alpha\gamma,{\cal G}_{\rm d}}(0) $ appear. After adding the 
simple contribution of ${\cal G}_{\rm D}$ computed from the Debye formula~(\ref{DebyeCC}), 
we eventually obtain the diagrammatic representation of $I_4$,
\be
\la{I4Mayer}
I_4= -\frac{1}{4\pi\beta \kappa_{\rm D}^2 } + \frac{1}{\kappa_{\rm D}^4} 
\sum_{{\cal G}_{\rm d}} \frac{1}{S_{{\cal G}_{\rm d}}}
\sum_{\alpha,\gamma} \rho_\alpha\rho_\gamma q_{\alpha} q_\gamma 
\tilde{K}_{\alpha\gamma,{\cal G}_{\rm d}}(0) \; .
\end{equation}

Representation~(\ref{I4Mayer}) is well-suited for computing the low-density expansion  of 
$I_4$. Indeed, and as usual, because of the density weights carried by the black points, 
only a finite number of graphs contribute up to a given order. However, here we have to take care 
of the dependence on the density of the bonds $b_{\rm D}$ and $b_{\rm R}$ through the Debye 
wavenumber $\kappa_{\rm D}=(\sum_\alpha 4 \pi \beta q_\alpha^2 \rho_\alpha)^{1/2}$. Consequently, the order 
of a contribution is not merely given by counting the number of black points on the one hand, while 
half-integer powers and logarithmic terms arise in the expansion on the other hand. We have computed 
the first three terms of that expansion, up to constant terms of order $\rho^0$ included. In the Appendix, 
we provide some technical details, as well as the complete list of graphs ${\cal G}_{\rm d}$ 
which contribute up to the considered order. The resulting expansion reads
\begin{multline}
\la{I4LDE}
I_4 = -\frac{1}{4\pi\beta \kappa_{\rm D}^2 } + 
\frac{\pi \beta^2}{\kappa_{\rm D}^5} \sum_{\alpha,\gamma} q_\alpha^3 q_\gamma^3 \rho_\alpha \rho_\gamma 
+\frac{2\pi\beta^3}{3 \kappa_{\rm D}^4} \sum_{\alpha,\gamma} q_\alpha^4 q_\gamma^4 \rho_\alpha \rho_\gamma
\ln(8\kappa_{\rm D}d_{\alpha\gamma}) \\
+\frac{4\pi\beta}{\kappa_{\rm D}^4} \sum_{\alpha,\gamma} q_\alpha^2 q_\gamma^2 \rho_\alpha \rho_\gamma d_{\alpha\gamma}^2
-\frac{3\pi\beta^2}{\kappa_{\rm D}^4} \sum_{\alpha,\gamma} q_\alpha^3 q_\gamma^3 \rho_\alpha \rho_\gamma d_{\alpha\gamma} \\
+\frac{1}{\kappa_{\rm D}^4} \sum_{\alpha,\gamma} q_\alpha q_\gamma \rho_\alpha \rho_\gamma \int \dd \br 
\left[\exp(-\beta u_{\alpha\gamma})-1+\beta u_{\alpha\gamma}-\beta^2 u_{\alpha\gamma}^2/2+\beta^3u_{\alpha\gamma}^3/6 \right] \\
+ \frac{11 \pi^2 \beta^4}{3 \kappa_{\rm D}^6} \sum_{\alpha,\gamma,\delta} q_\alpha^3 q_\gamma^3 q_\delta^4
\rho_\alpha \rho_\gamma \rho_\delta 
-\frac{52 \pi^3 \beta^5}{9\kappa_{\rm D}^8} \sum_{\alpha,\gamma,\delta,\eta} q_\alpha^3 q_\gamma^3 q_\delta^3 q_\eta^3
\rho_\alpha \rho_\gamma \rho_\delta \rho_\eta + o(\rho^0) \; .
\end{multline}

Not surprisingly, the leading term in the expansion~(\ref{I4LDE}) is the purely Debye contribution, and it behaves 
as $1/\rho$. The next correction is of order $1/\rho^{1/2}$ and is independent of the short-range part of the interactions. 
Those short-range parts arise in further corrections of order $\ln \rho$ and $\rho^0$. The last two terms of order $\rho^0$
are purely Coulomb contributions. The expansion is valid for any set of densities $(\rho_1,\rho_2)$, including of course the 
neutral sets defining the neutral TCP. For other short-range interactions, like hard cores for instance, the structure 
of the low-density expansion of $I_4$ is identical to that~(\ref{I4LDE}) explicitly computed for charged soft spheres.

\subsection{Free energy density}

The excess free energy of ${\cal S}^\ast$ for any set of homogeneous densities, can be obtained through the usual integration 
over the inverse temperature of the equilibrium average of the potential part of Hamiltonian~(\ref{HTCPB}). The resulting 
free energy density in thermal units $f^\ast(\rho_1,\rho_2,\beta)$ reduces to 
\be
\la{FEint}
f^\ast(\rho_1,\rho_2,\beta) = f_{\rm id}(\rho_1,\rho_2,\beta) + \frac{1}{2} \sum_{\alpha,\gamma} 
\int_0^\beta \dd \tau \int \dd \br \rho_{\alpha\gamma,\rm T}^\ast(r) u_{\alpha\gamma}(r) 
+ \frac{\beta}{2} \sum_{\alpha,\gamma} \rho_\alpha \rho_\gamma
\int \dd \br u_{\alpha\gamma}^{\rm SR}(r) \; .
\end{equation}
In that formula, the first term takes the familiar  form
\be
\la{FEideal}
f_{\rm id}(\rho_1,\rho_2,\beta) = \rho_1 \left[ \ln (\rho_1 (2 \pi \lambda_1^2)^{3/2}) -1 \right]
+ \rho_2 \left[ \ln (\rho_2 (2 \pi \lambda_2^2)^{3/2}) -1 \right]
\end{equation}
which describes a mixture of ideal gases. The next two terms account for interactions. 
The second term involving pair correlations $\rho_{\alpha\gamma,\rm T}^\ast(r)$ is obtained by adding 
and substracting  $u_{\alpha\gamma}^{\rm SR}$ to the purely Coulomb interactions in the particle-backround and 
background-background parts of the potential energy of Hamiltonian~(\ref{HTCPB}). This also provides the third term
which merely reduces to
\be
\la{FEbackground}
\frac{\beta}{2} \sum_{\alpha,\gamma} \rho_\alpha \rho_\gamma
\int \dd \br u_{\alpha\gamma}^{\rm SR}(r)= -2 \pi \beta \sum_{\alpha,\gamma} 
q_\alpha q_\gamma \rho_\alpha \rho_\gamma d_{\alpha\gamma}^2 \; .
\end{equation}
It is implicitly understood that pair correlations $\rho_{\alpha\gamma,\rm T}^\ast(r)$ in formula~(\ref{FEint}) are 
evaluated at inverse temperature $\tau$.  

The contribution of pair correlations $\rho_{\alpha\gamma,\rm T}^\ast(r)$ to $f^\ast$ follows by inserting 
its resummed diagrammatic representation~(\ref{MayerResummed}) into the second term of formula~(\ref{FEint}). The contribution 
of a given diagram ${\cal G}$ reads
\be
\la{FEgraph} 
\frac{1}{2 S_{\cal G}} \sum_{\alpha,\gamma} \rho_\alpha \rho_\gamma 
\int_0^\beta \dd \tau \int \dd \br \; u_{\alpha\gamma}(r) K_{\alpha\gamma,{\cal G}}^{(\tau)}(r)
\end{equation}
where $K_{\alpha\gamma,{\cal G}}^{(\tau)}(r)$ is the integral~(\ref{IntBlack}) over black points with ${\cal G}$ in place 
of ${\cal G}_{\rm d}$ and bonds evaluated at inverse temperature $\tau$. In the low-density limit, up to a given order 
in the density, only a finite number of 
contributions~(\ref{FEgraph}) needs to be retained. In the Appendix, we provide a few technical details of the calculations, as 
well as the list of graphs ${\cal G}$ which contribute to $f^\ast$ up to order $\rho^2$ included. Adding the simple 
ideal~(\ref{FEideal}) and background~(\ref{FEbackground}) contributions, we eventually obtain the low-density 
expansion of $f^\ast$,  
\begin{multline}
\la{FELDE}
f^\ast = \rho_1 \left[ \ln (\rho_1 (2 \pi \lambda_1^2)^{3/2}) -1 \right]
+ \rho_2 \left[ \ln (\rho_2 (2 \pi \lambda_2^2)^{3/2}) -1 \right] -\frac{\kappa_{\rm D}^3}{12 \pi} \\
-2 \pi \beta \sum_{\alpha,\gamma} q_\alpha q_\gamma \rho_\alpha \rho_\gamma d_{\alpha\gamma}^2 
+ \frac{3\pi\beta^2}{2} \sum_{\alpha,\gamma} q_\alpha^2 q_\gamma^2 \rho_\alpha \rho_\gamma d_{\alpha\gamma} 
-\frac{\pi\beta^3}{3} \sum_{\alpha,\gamma} q_\alpha^3 q_\gamma^3 \rho_\alpha \rho_\gamma
\ln(8\kappa_{\rm D}d_{\alpha\gamma}) \\
-\frac{1}{2} \sum_{\alpha,\gamma} \rho_\alpha \rho_\gamma \int \dd \br 
\left[\exp(-\beta u_{\alpha\gamma})-1+\beta u_{\alpha\gamma}-\beta^2 u_{\alpha\gamma}^2/2+\beta^3u_{\alpha\gamma}^3/6 \right] 
+ o(\rho^2) \; .
\end{multline}

The leading terms of order $\rho \ln \rho$ in the expansion~(\ref{FELDE}) are ideal contributions. The next correction of 
order $\rho^{3/2}$ arises 
from pure Coulomb interactions, and is nothing but the well-known Debye term. Contributions from the short range part 
of the interactions appear in the terms of order  $\rho^2 \ln \rho$ and  $\rho^2$. The terms left over 
are least of order  $\rho^{5/2} \ln \rho$. Expansion~(\ref{FELDE}) is valid for any set of densities $(\rho_1,\rho_2)$, 
and it gives access to all the other thermodynamical functions of ${\cal S}^\ast$ through suitable partial derivatives
with respect to  the independent parameters $\beta$, $\rho_1$ or $\rho_2$ defining an homogeneous equilibrium state of 
${\cal S}^\ast$. 

For other short-range regularizations, the low-density expansion of $f^\ast$ has the same structure 
as~(\ref{FELDE}). However, notice that for hard core potentials, the ideal term in the  
decomposition~(\ref{FEint}) of the corresponding $f^\ast$ must be replaced by the free energy density of 
hard spheres, $f_{\rm HS}(\rho_1,\rho_2,\beta)$. In the low-density limit, $f_{\rm HS}(\rho_1,\rho_2,\beta)$  
can be expanded in entire powers of $\rho$ around the ideal term~(\ref{FEideal}). In the resulting full expansion 
of $f^\ast$, there are terms which depend only on the hard core diameters 
$\sigma_{\alpha\gamma}$ and not on the particles charges.

\subsection{Checking the sum rule at lowest orders}

In order to check the sum rule~(\ref{SR4}), we first have to compute the low-density expansion of the partial 
compressibilities $\chi_{\alpha\gamma}^{-1}= \partial^2 f^\ast/\partial \rho_\alpha \partial \rho_\gamma$. 
Using expansion~(\ref{FELDE}) of $f^\ast$, we find
\begin{multline}
\la{chi11LDE}
\chi_{11}^{-1} = \frac{1}{\rho_1} - \frac{\pi \beta^2 q_1^4}{\kappa_{\rm D}} + 3 \pi \beta^2 q_1^4 d_{11}
- 4 \pi \beta q_1^2 d_{11}^2 
-\frac{2 \pi \beta^3 q_1^6}{3} \ln(8\kappa_{\rm D}d_{11}) \\
- \frac{8\pi^2 \beta^4 q_1^5}{3\kappa_{\rm D}^2} \sum_\alpha q_\alpha^3 \rho_\alpha 
+\frac{8\pi^3 \beta^5 q_1^4}{3\kappa_{\rm D}^4} \sum_{\alpha\gamma} q_\alpha^3 q_\gamma^3 \rho_\alpha \rho_\gamma \\
-\int \dd \br 
\left[\exp(-\beta u_{11})-1+\beta u_{11}-\beta^2 u_{11}^2/2+\beta^3u_{11}^3/6 \right]
+ o(\rho^0) \; ,
\end{multline}
\begin{multline}
\la{chi22LDE}
\chi_{22}^{-1} = \frac{1}{\rho_2} - \frac{\pi \beta^2 q_2^4}{\kappa_{\rm D}} + 3 \pi \beta^2 q_2^4 d_{22}
- 4 \pi \beta q_2^2 d_{22}^2 
-\frac{2 \pi \beta^3 q_2^6}{3} \ln(8\kappa_{\rm D}d_{22}) \\
- \frac{8\pi^2 \beta^4 q_2^5}{3\kappa_{\rm D}^2} \sum_\alpha q_\alpha^3 \rho_\alpha 
+\frac{8\pi^3 \beta^5 q_2^4}{3\kappa_{\rm D}^4} \sum_{\alpha\gamma} q_\alpha^3 q_\gamma^3 \rho_\alpha \rho_\gamma \\
-\int \dd \br 
\left[\exp(-\beta u_{22})-1+\beta u_{22}-\beta^2 u_{22}^2/2+\beta^3u_{22}^3/6 \right]
+ o(\rho^0) \; ,
\end{multline}
and
\begin{multline}
\la{chi12LDE}
\chi_{12}^{-1} = \chi_{21}^{-1} =  - \frac{\pi \beta^2 q_1^2 q_2^2}{\kappa_{\rm D}} + 3 \pi \beta^2 q_1^2 q_2^2 d_{12}
- 4 \pi \beta q_1 q_2 d_{12}^2 
-\frac{2 \pi \beta^3 q_1^3 q_2^3}{3} \ln(8\kappa_{\rm D}d_{12}) \\ 
-\frac{ \pi \beta^3 q_1^3 q_2^3}{3}
- \frac{4\pi^2 \beta^4 q_1^2 q_2^2}{3\kappa_{\rm D}^2} \sum_\alpha q_\alpha^4 \rho_\alpha 
+\frac{8\pi^3 \beta^5 q_1^2 q_2^2}{3\kappa_{\rm D}^4} \sum_{\alpha\gamma} q_\alpha^3 q_\gamma^3 \rho_\alpha \rho_\gamma\\
-\int \dd \br 
\left[\exp(-\beta u_{12})-1+\beta u_{12}-\beta^2 u_{12}^2/2+\beta^3u_{12}^3/6 \right]
+ o(\rho^0) \; .
\end{multline}
Notice that the leading contributions of order $1/\rho$ in both 
$\chi_{11}^{-1}$ and $\chi_{22}^{-1}$ arise from the ideal terms in 
$f^\ast$, while the next correction of order $1/\rho^{1/2}$ comes from the Debye term in expansion~(\ref{FELDE}).
The leading contribution of order $1/\rho^{1/2}$ in $\chi_{12}^{-1}$ is also provided by that Debye correction. 
All terms which are left over in expansions~(\ref{chi11LDE}), (\ref{chi22LDE}) and (\ref{chi12LDE}) are at least 
of order $\rho^{1/2} \ln \rho$. 

According to the expression~(\ref{IsoCompbis}) of the isothermal compressibility, the 
thermodynamical quantity in the right hand side of sum rule~(\ref{SR4}) can be rewritten as 
\be
-\frac{\rho^2 }{(4\pi (q_1-q_2))^2 \beta^3} (\chi_{11}^{-1}\chi_{22}^{-1}-\chi_{12}^{-2}) \chi_T
= -\frac{(\chi_{11}^{-1}\chi_{22}^{-1}-\chi_{12}^{-2}) }
{(4\pi \beta)^2 (q_2^2\chi_{11}^{-1} + q_1^2\chi_{22}^{-1}-2 q_1 q_2\chi_{12}^{-1})} \; .
\la{SR4thermo}
\end{equation}  
The low-density expansion of that thermodynamical expression is straightforwardly computed by using the 
expansions~(\ref{chi11LDE}), (\ref{chi22LDE}) and (\ref{chi12LDE}) of the $\chi_{\alpha\gamma}^{-1}$'s. 
Its leading behaviour is immediately obtained by noticing that
both $\chi_{11}^{-1}$ and $\chi_{22}^{-1}$ diverge faster than $\chi_{12}^{-1}$ in the zero-density limit, and the 
corresponding purely ideal behaviours $\chi_{11}^{-1} \sim 1/\rho_1$ and  $\chi_{22}^{-1} \sim 1/\rho_2$ provide  
\be
-\frac{(\chi_{11}^{-1}\chi_{22}^{-1}-\chi_{12}^{-2}) }
{(4\pi \beta)^2 (q_2^2\chi_{11}^{-1} + q_1^2\chi_{22}^{-1}-2 q_1 q_2\chi_{12}^{-1})}
\sim -\frac{\rho_1^{-1} \rho_2^{-1}}{(4\pi \beta)^2 (q_2^2 \rho_1^{-1} + q_1^1 \rho_2^{-1} )} =  
-\frac{1}{4\pi\beta \kappa_{\rm D}^2 }  \; ,
\la{SR4thermoZD}
\end{equation}  
which coincides with the leading term in expansion~(\ref{I4LDE}) of $I_4$. The calculation of the next correction  
of order $1/\rho^{1/2}$ remains simple, since it requires to retain only the first 
Debye corrections of order  $1/\rho^{1/2}$ to the ideal terms in both $\chi_{11}^{-1}$ and $\chi_{22}^{-1}$, while 
$\chi_{12}^{-1}$ can be replaced by its leading Debye behaviour. The determination of the terms of order 
$\rho^0$ and $\rho^0 \ln \rho$ is still straightforward but more cumbersome. Eventually, we find that all those corrections   
to the ideal behaviour~(\ref{SR4thermoZD}) of the thermodynamical quantity~(\ref{SR4thermo}) 
exactly match the low-density expansion~(\ref{I4LDE}) of $I_4$ inferred from its microscopic definition. Thus, the fourth 
moment sum rule perfectly works, at least up to the considered order in the density.

\section{Concluding comments and perspectives} 
\la{Conclusion}

In this paper, we have derived a new sum rule for the fourth moment of charge correlations of a TCP. 
Since the Stillinger-Lovett second moment sum rule naturally emerges as a by-product of our analysis, 
we believe that this new sum rule holds in any conducting phase, although 
all the steps of its derivation are not under a complete mathematical control at the moment. In particular, 
we expect that the free energy functional can be safely 
expanded around homogeneous states inside the conducting phase. We stress that all the partial compressibilities 
of the auxiliary system, namely second partial derivatives with respect 
to particle densities of the free energy density,
have then to be well defined. Thus critical points must be dealt with some care, since 
singularities in the thermodynamical quantities arise on the one hand, while perfect screening properties can be lost 
on the other hand, as mentioned below. 

Our derivation also involves implicit assumptions about the existence of the thermodynamic limit, 
and of intrinsic bulk properties with bulk densities which become homogeneous far from the boundaries. 
Strictly speaking, to our knowledge, this has been only 
proved for the general three-dimensional TCP in the Debye regime~\cite{BryFed1980} and for its charge-symmetric 
version~\cite{FroPar1978}. In two dimensions, 
where the Coulomb potential takes a logarithmic form, both the neutrality and homogeneity of 
a TCP of point charges have been proved~\cite{Serfaty}. Extensions of such results to all the systems introduced here 
would be quite valuable of course, and might constitute the first steps towards 
a complete proof of our sum rule. Meanwhile, physical arguments, 
in particular related to the beautiful proof for quantum Coulomb matter~\cite{LL72}, 
strongly suggest that the classical TCP, as well as its 
version immersed in a charged uniform background, do
sustain a well-behaved TL. Furthermore, there are strong evidences, 
arising either from specific models or mean-field approaches, 
that screening properties in the bulk 
can be disentangled from the reorganization of charges at the surface, so any boundary effects can be indeed 
\textit{a priori} ignored. 

In the absence of a complete mathematical proof, checking the sum rule within exact calculations for 
specific models or thermodynamical regimes is particularly valuable. Here, such checking has been carried out 
for charged soft spheres in the low density regime, through the explicit calculation of the lowest order terms in density 
expansions of the quantities of interest. This illustrates the subtle interplay between short-range and 
screened Coulomb contributions which ultimately ensure the validity of the fourth moment sum rule at the considered orders. 
If there exists a simple reorganization of the Abe-Meeron diagrams which shows the validity of the second moment 
Stillinger Lovett sum rule at any order in the density expansion, a similar trick  
for the fourth moment sum rule, certainly more cumbersome, remains to be discovered. 

If our derivation of the fourth moment sum rule is based on the response to external perturbations, 
more intrinsic derivations would be of great interest, both for enforcing its expected validity on the 
one hand, and for sheding light on the internal mechanisms at work on the other hand. For instance, the second moment 
Stillinger Lovett sum rule can be retrieved within suitable manipulations of the BGY hierarchy 
equations for the equilibrium distribution functions of the unperturbed homogeneous TCP, as 
shown by Gruber and Martin~\cite{GruMar1983}. Moreover, the BGY hierarchy 
equations have been also used for deriving the fourth moment sum rule for the 
OCP~\cite{BraVieille} and for the BIM with pure Coulomb interactions~\cite{Suttorp2008}. We are looking for 
extending such derivations to the TCP case, where the presence of short-range interactions requires 
further manipulations. Notice, that a full reorganization of Abe-Meeron diagrammatics as described above 
could be also seen as an intrinsic derivation. In the same spirit, let us mention that 
a sixth moment sum rule for the charge correlations of the two-dimensional OCP with logarithmic interactions
was established through a full term by term analysis of the Abe-Meeron diagrammatics~\cite{Kal}. 
 
The fourth moment sum rule obviously extends to the TCP immersed in a charged background, namely 
the thermodynamical expression~(\ref{SR4}) of the fourth moment of charge correlations is not restricted 
to densities satisfying overall neutrality, but it is valid for any set of densities $(\rho_1,\rho_2)$. This is well 
illustrated by the low density calculations for charged soft spheres. Moreover, a similar 
DFT analysis combined with linear response theory can be carried out for an arbitrary number $n$ of components. 
This would lead to formulae analogous to the thermodynamical expression~(\ref{SR4}), but with a more complicated structure
arising from the inversion of a $n \times n$ matrix. In the case of the MIM, they 
should be equivalent to those derived by Suttorp~\cite{Suttorp2008}. Eventually, within our 
formalism, one can obtain sum rules for the 
zeroth and second moment of particle correlations, by comparing the DFT calculation of a given particle density 
to its linear response expression. Such sum rules are again equivalent to those obtained for the MIM~\cite{Suttorp2008}. 

Among the various possible applications of our new sum rule, we would like to emphasize 
its usefulness for a better understanding of the plausible lack of screening properties at the 
ionic critical point. The liquid-gas transition of a TCP
has been widely studied the last twenty years. Let us mention for instance two 
recent works~\cite{DKF2012}, \cite{CaillolLevesque2014}. Numerical simulations have
convincingly shown that both liquid and gas phases 
display perfect screening properties, namely the second moment Stillinger-Lovett
sum rule is satisfied. However, a first suspicion about the violation of that sum rule 
at the critical point was pointed out by Caillol~\cite{Caillol1995}. Meanwhile, such violation was also 
observed for a solvable asymmetric mean-spherical model by Aqua and Fisher~\cite{AquaFisher2004}, which is expected to 
share common properties with an asymmetric TCP. More recently, and contrarily to various 
theoretical expectations, the violation of the Stillinger Lovett sum rule 
at the critical point was also observed for the fully symmetric RPM by 
Das, Kim and Fisher~\cite{DasKimFisher2011} : they provide strong numerical evidences 
by combining refined Monte Carlo simulations in the 
grand-canonical ensemble with finite-size scaling methods. Furthermore, they also 
show that the fourth moment of charge correlations diverges when approaching the critical point, 
in a way analogous to the isothermal compressibility. Clearly, our thermodynamical 
expression~(\ref{SR4}) of that fourth moment constitutes a promising tool for analyzing its behaviour
near the critical point, as well as the underlying coupling between charge and mass fluctuations.   

Eventually, let us conclude by a few comments regarding the two dimensional case. In two 
dimensions (2D), the Coulomb potential, defined as the solution of Poisson equation, 
takes the well-known logarithmic form. Thanks to the relatively weak singularity at the origin 
of the 2D Coulomb potential, the TCP of point charges is well behaved for coupling constants $\Gamma < 2$, namely 
at sufficiently high temperatures~\cite{DeuLav}. However, short-range interactions need to be introduced for $\Gamma \geq 2$ 
in order to avoid the collapse between positive and negative charges. The fourth moment sum rule derived here 
explicitly in three dimensions (3D), can be straightforwardly extended to the 2D case : this leads to the simple 
replacement of the factor $4 \pi$ in formula~(\ref{SR4}) by the factor $2 \pi$, a direct consequence 
of the modification of Poisson equation when changing from 3D to 2D. At $\Gamma = 2$, 
Cornu and Jancovici~\cite{CorJanco1988} exploited a mapping with a field theory model valid 
for pure Coulomb interactions, which allowed them to derive analytical expressions 
for particle correlations of the 2D TCP. Such expressions are expected to become exact in the 
zero density limit for the well-behaved TCP including short-range interactions. Thus, they constitute 
a reliable starting point for further checking of the fourth moment sum rule 
at finite coupling. Similarly to its application to the study of ionic criticality in 3D, the 
fourth moment sum rule should also bring new insights for the celebrated 
Kosterlitz-Thouless transition~\cite{KosThou1973}, \cite{Minhagen} : in the temperature-density plane, there 
appears a line of critical points separating a high-temperature conducting phase from a low-temperature 
dielectric phase~\cite{AlaCor1997}. The implications of the fourth moment sum rule 
should complete the results of a previous work~\cite{AlaCor1992}, where a plausible scenario
for the large-distance decay of particle correlations in the dielectric phase was constructed in 
a way consistent with various sum rules.       

\textbf{ACKOWLEDGMENT} This work was presented at the conference in the memory of Bernard Jancovici 
(Institut Henri Poincar\'e, Paris, 5 and 6 november 2015), and it is dedicated to him. We would also like to thank 
Michael Fisher for his stimulating interest and useful discussions. 

\textbf{APPENDIX}

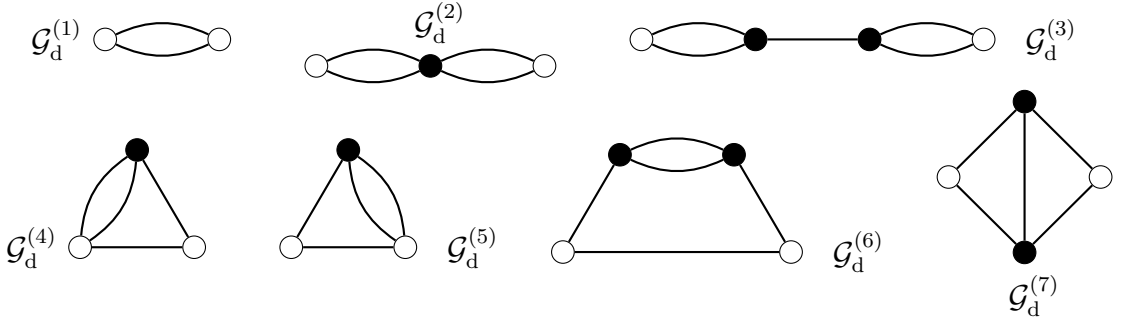
\begin{figure} 
	\begin{center}
		\begin{tikzpicture}
			\node[blanc] (R0) at (0,0) [label=left: \; ${\cal G}_{\rm d}^{(1)}$ ]{};
			\node[blanc] (R1) at (1.5,0) {};
			\drawLinkFR{R0}{R1};
		\end{tikzpicture}
		\quad\quad
		\begin{tikzpicture}
			\node[blanc] (R0) at (0,0) {};
			\node[noir] (R1) at (1.5,0) [label=above: \; ${\cal G}_{\rm d}^{(2)}$ ]{};
			\drawLinkFR{R0}{R1};
			\node[blanc] (R2) at (3,0) {};
			\drawLinkFR{R1}{R2}
		\end{tikzpicture}
		\quad\quad
		\begin{tikzpicture}
			\node[blanc] (R0) at (0,0) {};
			\node[noir] (R1) at (1.5,0) {};
			\drawLinkFR{R0}{R1};
			\node[noir] (R2) at (3,0) {};
			\drawLinkFC{R1}{R2};
			\node[blanc] (R3) at (4.5,0) [label=right: \; ${\cal G}_{\rm d}^{(3)}$ ]{};
			\drawLinkFR{R2}{R3}
		\end{tikzpicture}
		\\
		\begin{tikzpicture}
			\node[blanc] (R0) at (0,0) [label=left: \; ${\cal G}_{\rm d}^{(4)}$ ]{};
			\node[noir] (R1) at (0.75,1.29) {};
			\drawLinkFR{R0}{R1};
			\node[blanc] (R2) at (1.5,0) {};
			\drawLinkFC{R1}{R2};
			\drawLinkFC{R0}{R2};
		\end{tikzpicture}
		\quad\quad
		\begin{tikzpicture}
			\node[blanc] (R0) at (0,0) {};
			\node[noir] (R1) at (0.75,1.29) {};
			\drawLinkFC{R0}{R1};
			\node[blanc] (R2) at (1.5,0) [label=right: \; ${\cal G}_{\rm d}^{(5)}$ ] {};
			\drawLinkFR{R1}{R2};
			\drawLinkFC{R0}{R2};
		\end{tikzpicture} 
		\quad
		\begin{tikzpicture}
			\node[blanc] (R0) at (0,0) {};
			\node[noir] (R1) at (0.75,1.29) {};
			\drawLinkFC{R0}{R1};
			\node[noir] (R2) at (2.25,1.29) {};
			\drawLinkFR{R1}{R2};
			\node[blanc] (R3) at (3,0) [label=right: \; ${\cal G}_{\rm d}^{(6)}$ ] {};
			\drawLinkFC{R2}{R3};
			\drawLinkFC{R0}{R3};
		\end{tikzpicture}
		\quad
		\begin{tikzpicture}
			\node[blanc] (R0) at (0,0) {};
			\node[noir] (R1) at (1,1) {};
			\drawLinkFC{R0}{R1};
			\node[blanc] (R2) at (2,0) {};
			\drawLinkFC{R1}{R2};
			\node[noir] (R3) at (1,-1) [label=below: \; ${\cal G}_{\rm d}^{(7)}$ ]{};
			\drawLinkFC{R0}{R3};
			\drawLinkFC{R3}{R2};
			\drawLinkFC{R1}{R3};
		\end{tikzpicture} 
	\end{center}
	\caption{The seven diagrams ${\cal G}_{\rm d}$ which contribute to $I_4$ up to order $\rho^0$ included 
	in the diagrammatic series~(\ref{I4Mayer}). The bubbles represent short-range dressed bonds $b_{\rm R}$~(\ref{BondSRD}).} 
	\label{I4diagrams}
\end{figure}

In Fig.~\ref{I4diagrams}, we list the seven diagrams ${\cal G}_{\rm d}$ 
which contribute to the diagrammatic series~(\ref{I4Mayer})
for the fourth moment $I_4$ up to order $\rho^0$ included.
When computing the Fourier transform
$\tilde{K}_{\alpha\gamma,{\cal G}_{\rm d}}(0)$ for each of those diagrams, we can apply the convolution theorem 
at various places, namely with intermediate points which reduce either to the black points for graphs 
${\cal G}_{\rm d}^{(2)}$ and ${\cal G}_{\rm d}^{(3)}$, or to the root white points 
for graphs ${\cal G}_{\rm d}^{(4-7)}$ by exploiting translational 
invariance. Two key quantities turn then to be the inverse Fourier transform of $[\tilde{\phi}_{\rm D} (k)]^2$ and 
$[\tilde{\phi}_{\rm D} (k)]^3$ which reduce respectively to
\begin{equation}
\frac{1}{(2\pi)^3} \int \dd \bk \exp(-i \bk \cdot \br) \frac{16 \pi^2}{(k^2 + \kappa_{\rm D}^2)^2} 
=\frac{2 \pi}{\kappa_{\rm D}} \exp(-\kappa_{\rm D} r)
\label{A1}
\end{equation}
and 
\begin{equation}
\frac{1}{(2\pi)^3} \int \dd \bk \exp(-i \bk \cdot \br) \frac{64 \pi^3}{(k^2 + \kappa_{\rm D}^2)^3} 
=\frac{2 \pi^2}{\kappa_{\rm D}^3} (1+\kappa_{\rm D} r ) \exp(-\kappa_{\rm D} r) \; ,
\label{A1bis}
\end{equation}
after a straightforward application of the theorem of residues. Another useful trick relies on the decomposition
\begin{equation}
b_{\rm R} = b_{\rm R}^{(\rm T)} -\beta u_{\alpha\gamma}^{\rm SR} + 
\frac{\beta^2}{2} (u_{\alpha\gamma}^{\rm SR} + q_\alpha q_\gamma \phi_{\rm D})^2 -
\frac{\beta^3}{6} (u_{\alpha\gamma}^{\rm SR} + q_\alpha q_\gamma \phi_{\rm D})^3 
\label{A2}
\end{equation}
with the truncated bond
\begin{equation}
b_{\rm R}^{(\rm T)} = \exp (-\beta (u_{\alpha\gamma}^{\rm SR} + q_\alpha q_\gamma \phi_{\rm D})) -1 
+\beta (u_{\alpha\gamma}^{\rm SR} + q_\alpha q_\gamma \phi_{\rm D}) 
-\frac{\beta^2}{2} (u_{\alpha\gamma}^{\rm SR} + q_\alpha q_\gamma \phi_{\rm D})^2 
+\frac{\beta^3}{6} (u_{\alpha\gamma}^{\rm SR} + q_\alpha q_\gamma \phi_{\rm D})^3 \; .
\label{A3}
\end{equation}
Indeed, the corresponding
contribution of the truncated bond $b_{\rm R}^{(\rm T)}$ in graphs ${\cal G}_{\rm d}^{(1-5)}$ can be computed 
at lowest order in the density by merely replacing $u_{\alpha\gamma}^{\rm SR} + q_\alpha q_\gamma \phi_{\rm D}$
by the bare pair potential $u_{\alpha\gamma}$ since $(\exp (-\beta u_{\alpha\gamma}) -1 
+\beta u_{\alpha\gamma} -\beta^2 u_{\alpha\gamma}^2/2 + \beta^3 u_{\alpha\gamma}^3/6) $ is integrable in the whole space. The 
next density-dependent corrections to that leading contribution behave as
$\rho^{1/2}$ and can thus be neglected in the considered calculation of $I_4$ up to order $\rho^0$. The  
contributions of the other terms in the decomposition~(\ref{A2}) are easily computed thanks to the simple 
analytic expressions of $u_{\alpha\gamma}^{\rm SR}$ and $\phi_{\rm D}$. Eventually, combining the above convolution and 
decomposition tricks, we obtain formula~(\ref{I4LDE}) for $I_4$. 

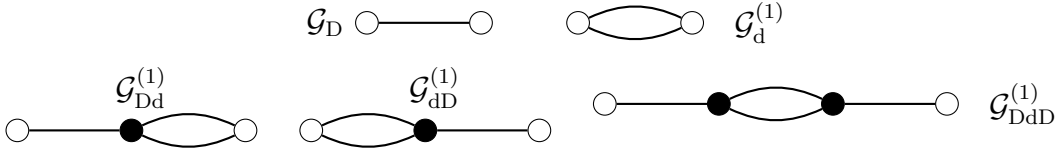
\begin{figure} 
	\begin{center}
	  \begin{tikzpicture}
			\node[blanc] (R0) at (0,0) [label=left: \; ${\cal G}_{\rm D}$ ]{};
			\node[blanc] (R1) at (1.5,0) {};
			\drawLinkFC{R0}{R1};
		\end{tikzpicture}
		\quad\quad
		\begin{tikzpicture}
			\node[blanc] (R0) at (0,0) {};
			\node[blanc] (R1) at (1.5,0) [label=right: \; ${\cal G}_{\rm d}^{(1)}$ ] {};
			\drawLinkFR{R0}{R1};
		\end{tikzpicture}
		\\ 
		\begin{tikzpicture}
			\node[blanc] (R0) at (0,0) {};
			\node[noir] (R1) at (1.5,0) [label=above: \; ${\cal G}_{\rm Dd}^{(1)}$ ]{};
			\drawLinkFC{R0}{R1};
			\node[blanc] (R2) at (3,0) {};
			\drawLinkFR{R1}{R2};
		\end{tikzpicture}
		\quad
		\begin{tikzpicture}
			\node[blanc] (R0) at (0,0) {};
			\node[noir] (R1) at (1.5,0) [label=above: \; ${\cal G}_{\rm dD}^{(1)}$ ]{};
			\drawLinkFR{R0}{R1};
			\node[blanc] (R2) at (3,0) {};
			\drawLinkFC{R1}{R2};
		\end{tikzpicture}
		\quad
		\begin{tikzpicture}
			\node[blanc] (R0) at (0,0) {};
			\node[noir] (R1) at (1.5,0) {};
			\drawLinkFC{R0}{R1};
			\node[noir] (R2) at (3,0) {};
			\drawLinkFR{R1}{R2};
			\node[blanc] (R3) at (4.5,0) [label=right: \; ${\cal G}_{\rm DdD}^{(1)}$ ] {};
			\drawLinkFC{R2}{R3};
		\end{tikzpicture}
	\end{center}
	\caption{The five diagrams ${\cal G}$ which contribute to $f^\ast$ up to order $\rho^2$ included. } 
	\label{FEdiagrams}
\end{figure}

The five graphs in the series~(\ref{MayerResummed}) for particle correlations 
which provide contributions~(\ref{FEgraph}) to the free-energy density $f^\ast $ 
are listed in Fig.~\ref{FEdiagrams}. 
Each contribution follows from formula~(\ref{FEgraph}), so the value 
$K_{\alpha\gamma,{\cal G}}^{(\tau)}(r)$ of each graph is first computed with
bonds $b_{\rm D}$ and $b_{\rm R}$ evaluated at temperature $\tau$. After multiplication of 
$K_{\alpha\gamma,{\cal G}}^{(\tau)}(r)$  by 
the pair potential $u_{\alpha\gamma}(r)$, the further integrals over $\br$ in the whole space 
are readily computed by using decomposition~(\ref{A2}) with $\tau$ in place of $\beta$, as well as the 
inverse Fourier transforms of $[\tilde{\phi}_{\rm D} (k)]^2$ given by expression~(\ref{A1}), 
and of $\tilde{\phi}_{\rm D} (k) 4 \pi/k^2$ which reduces to
\begin{equation}
\frac{1}{(2\pi)^3} \int \dd \bk \exp(-i \bk \cdot \br) \frac{16 \pi^2}{k^2 (k^2 + \kappa_{\rm D}^2)} 
=\frac{4 \pi}{\kappa_{\rm D}^2 r} (1-\exp(-\kappa_{\rm D} r)) \; . 
\label{A4}
\end{equation}
The final integrals over $\tau$ from $0$ to $\beta$ are then easily and explicitly performed for 
all terms which reduce to combinations of powers laws and logarithmic terms. It remains a term involving 
\begin{equation}
\int \dd \br \; u_{\alpha\gamma} \; [ \exp (-\tau u_{\alpha\gamma}) -1 
+\tau u_{\alpha\gamma} 
-\frac{\tau ^2}{2} u_{\alpha\gamma}^2 ] \; ,
\label{A5}
\end{equation}
whose integration over $\tau$ leads to the last correction of order $\rho^2$ in the 
formula~(\ref{FELDE}) for $f^\ast$.

\end{document}